\documentclass[aps,prb,onecolumn,superscriptaddress,10pt]{revtex4-2}
\pdfoutput=1 
\usepackage[utf8]{inputenc}
\usepackage[english]{babel}
\usepackage[T1]{fontenc}
\usepackage[pdftex]{graphicx}
\usepackage{amsmath}
\usepackage{cancel}
\usepackage{amsthm}
\usepackage{amssymb}
\usepackage{braket}
\usepackage{xcolor}
\usepackage{dcolumn}
\usepackage{bbm}
\usepackage{bbold}
\usepackage[normalem]{ulem}
\usepackage{footmisc}
\usepackage{xcolor}
\usepackage{booktabs}
\usepackage{comment}
\usepackage{multirow}
\usepackage[colorlinks=true,linkcolor=blue,citecolor=blue,urlcolor=blue,unicode]{hyperref}

\usepackage{amsmath}
\usepackage{bm}
\usepackage{graphicx}
\usepackage{subfigure}
\usepackage{geometry}
\usepackage{ulem}
\usepackage{indentfirst}
\usepackage{amsfonts,amssymb,dsfont}
\usepackage{accents}

\usepackage{setspace}
\usepackage{threeparttable}
\usepackage{amsmath,mathtools,amsthm}
\usepackage{array}
\usepackage{extarrows}
\usepackage{upgreek}
\usepackage{orcidlink}

\newcommand{\m}[1]{\textcolor{black}{#1}}

\makeatletter
\renewcommand*\env@matrix[1][\arraystretch]{%
	\edef\arraystretch{#1}%
	\hskip -\arraycolsep
	\let\@ifnextchar\new@ifnextchar
	\array{*\c@MaxMatrixCols c}}
\makeatother

\usepackage{orcidlink}

\begin{document}
	\begin{small}
		\title{Local boson-nonlocal boson coupling in a four-level system:  Adiabatic, non-adiabatic, and non-Hermitian effects}		
		\author{Chen-Huan Wu 
		\orcidlink{0000-0003-1020-5977} }
			\thanks{chenhuanwu1@gmail.com}
		\affiliation{College of Physics and Electronic Engineering, Northwest Normal University, Lanzhou 730070, China}
		

		%

		\begin{abstract}
			
			We investigate the dynamics of an open quantum system comprising a two-level electronic system coupled to local boson mode and a bosonic bath. The system is described by four distinct states, including the ground and excited electronic states, each with its corresponding zero- and one-boson vibrational levels. The dissipative dynamics arising from interactions with an external environment are modeled using two distinct theoretical frameworks: the standard Lindblad master equation and a non-Hermitian effective Hamiltonian approach. We derive the full Liouvillian superoperator for both formalisms, revealing a crucial distinction: while the Lindblad equation accounts for both state decay and repopulation via quantum jumps, the non-Hermitian formalism only captures the decay, leading to non-conservation of the total system probability. 
		\end{abstract}

		\maketitle

		\section{Introduction}
		
		Understanding the dynamics of open quantum systems is of paramount importance in fields ranging from quantum chemistry and photophysics to quantum information science. A system's interaction with its surrounding environment gives rise to a range of phenomena including dissipation, decoherence, and relaxation. The standard theoretical tool for modeling such irreversible processes is the Lindblad master equation, which provides a rigorous and physically consistent description of the time evolution of the system's density matrix.

		In non-Hermitian case, the thermalization is allowed
		by the diffusive spreading of initial state information 
		(or correlations) over the entire system (upto long-time limit)
		and meanwhile small local subsystems are possible to loss the memory about 
		initial information and relax to thermal state under weak disorder.
		This corresponds to the non-local measurement, e.g., the measurement ideally affects all
		the bonds in a spin chain model,
		which will leads to superballistic spreading (of entanglement or quantum information)
		under nonunitary evolution\cite{Sriram}.
		Such superballistic spreading has a faster-than-linear entanglement growth,
		which appears in the initial stage of spreading\cite{Kim}
		where there is equivalent bipartition with maximal thermodynamic entropy.
		While in the presence of nonlocal measurement and unitary evolution,
		there is ballistic spreading with linear entanglement growth
		following the volume law.
		Distinct to the former,
		the ballistic spreading involving the process with information spread
		from one heavy carrier to another heavy carrier through the carrier-carrier interaction, 
		such that the entanglement is generated between those carriers
		(with information scrambling).
		An example of such carrier (a local degree of freedom) satisfying the volume scaling law 
		is the spin\cite{Kim,Cheng}.
		In contrast to nonlocal measurement, 
		the local measurement will not prohibit the global thermalization,
		whose impact is exponentially localized\cite{Chandran}.
		Importantly,
		the non-local measurement cause the ballistic entanglement spreading 
		(and diffusive entanglement spreading throughout the system in the late-time stage
		\cite{Cheng,Kim})
		in non-Hermitian case where the perturbation from
		local projective measurement is avoided by the effective single-particle picture,
		coexist with the diffusive energy (or information carriers) transport 
		throughout the whole system.
		Except the continuous level distribution\cite{Imbrie},
		we further found that the equally-spaced eigenvalues (of the thermal states)
		also reveals the invalidity of perturbation theory.
		In this case,
		the only local observable is the total energy,
		thus there is no ballistic transport in terms of ballistically traveling quasiparticles\cite{Kim},
		like photon is absent,
		and the only way for energy transport is diffusion (through strong scattering).

		In the many-body case,
		a large amount of bound states (mode) emerge discretely
		where the superpositional coherence is allowed between the open subsystems.
		Also,
		the discretely emergent non-Hermitian skin effect with unthermalized-type disentangling
		cause the absence of measurement-induced entanglement transition
		(volume-law to area-law)\cite{Wang}.
	Consequently, the resulting long-lived  mixed-state density matrices often maintain non-zero coherences despite being subjected to decoherence. 
	This is a signature of dissipation-induced entanglement or non-thermal steady states.
		\m{ Bound states similar to the bound states in the continuum (BIC) in non-Hermitian models are studied recently\cite{Kattel,Kattel2}.
		In this article, we found that a block-diagonal Liouvillian with multiple zero eigenvalues is available in both the Hermitian and non-Hermitian cases.}

		The weak disorder and the non-Hermiticity result in the delocalization in many-body case,
		which means the system is free from the constraints of integrability and many-body localization,
		and, in the mean time,
		without thermalization and global quantum correlation or entanglement\cite{Cheng}.
		The non-Hermiticity is realized by the constriction to a subspace in finite size
		(symmetry sector)
		with the conditioned (postselected) dynamic of quantum trajectories 
		on the measurement outcomes,
		where exponentially many measurements are required.
		The measurements on the conditoned dynamics in non-Hermitian system 
		return null results 
		due to the lacking of average over trajectory ensemble
		as well as the absence of randomness in the measured dynamics.
		non-Hermitian case always process a linearly-in-time (volume law) to saturating (area law) transition,
		many-body case (without postselection) process a volume-law entanglement without saturating\cite{Cipolloni}
		and exhibit persistent fluctuations with the increasing complexity of coherence configuration,
		i.e., the set of local Hamiltonian (disorder-free).
		It is such locality and coherence that guarantee the fidelity (as well as the subsystem purity).
		The suppression on fluctuation implies the common effect between the continuous strong dissipations
		and the on-site disorder which is capable to leads to localization even in non-Hermitian many-body system\cite{Wang2}.
		The diffusion-induced continuous dissipations here is nonlocal and related to the widely spreading quantum correlations. 
		Another property similar to the Zeno effect is the nonstabilizerness whose additivity guarantees its global nature
		and can even be produced by a unitary operator 
		by averaging over the ensemble of pure state trajectories of a full unitary group\cite{Fux,Leone}.
		
		The quantum Zeno effect is significant in later stage of non-Hermitian case,
		whereas in early stage the large system size can enhance the robustness of entanglement grow against the
		classical stochastic dynamics by the quantum jumps\cite{Le,Roscilde},
		i.e., lower the effect of dissipation.
		Note that the heating loss (suppression on entanglement production)
		here is different to that in the many-body case,
		but relies on the robust initial state property with non-Hermitian collapses and revivals\cite{Turkeshi},
		similar to the quantum cooling procotol of Ref.\cite{cooling} which is by reducing the excited
		states back to ground states.
		\m{The relationship between the Zeno effect and dissipation (dephasing noise) in many-body quantum systems is recently studied by Refs.\cite{Javed,Tang},
and it is found that, the dissipation does not simply freeze the dynamics, instead, it can either slow down (Zeno) or speed  up (anti-Zeno) the dynamics, depending on the system's parameters and the timescale of observation.}
		While
		the Loschmidt echo directly indicates that the discrete symmetries only exists in many-body case
		with the jump-induced local dissipations (accompanied by the local dephasing and oscillatory behavior\cite{Dolgirev})
		and dissipative coupling via the ballistic transports and synchronization,
		which is unique to a many-body system and representing the robust initial-state fidelity under the measurements at a finite rate\cite{Chan,Li,Zhang2,Znidaric}.
		Except the many-body nature,
		the infinitely long time needed to loss the initial-state information
		also consistent with the thermodynamic limit in many-body case,
		where there are thermal baths holding well-defined temperature.

		In this work, we focus on a specific open quantum system model that captures the essential physics of molecular transitions. The system consists of a two-level electronic manifold, with states $|0\rangle_e$ (ground) and $|1\rangle_e$ (excited), coupled to a single local boson mode representing a vibrational degree of freedom. This simplified model gives rise to a four-level system.
		The transitions between these states are fundamental to understanding phenomena such as the zero-phonon line (electronic transition without a change in vibrational state) and phonon sidebands (electronic transitions accompanied by a change in vibrational state). The initial state of the system is prepared in the ground vibronic state and, upon interaction with a light field, the initial vibrational wavepacket is "instantly projected" onto the potential energy surface of the excited electronic state, as dictated by the Franck-Condon principle\cite{Franck}. 
		
		We consider two distinct aspects to model the dissipative dynamics of this system. The first aspect utilizes the full Lindblad master equation, which precisely accounts for all coherent and incoherent processes, including the quantum jump terms responsible for transitions between energy levels. The second aspect employs a non-Hermitian effective Hamiltonian,
		with no-jump evolution and a non-trace-preserving Liouvillian. This method simplifies the problem by incorporating the dissipative effects directly into the Hamiltonian via a complex-valued term. 
		We explicitly construct the Liouvillian superoperator for both the full Lindblad equation and the non-Hermitian effective Hamiltonian. By analyzing these superoperators, we will rigorously demonstrate the mathematical and physical limitations of the non-Hermitian approach.
		Our results are helpful in modeling open quantum systems in molecular and quantum contexts.
		
		\section{model}
		We consider the system involving the interaction between local boson and nonlocal boson, which reads
		\begin{equation} 
			\begin{aligned}
				&		H=H_{s}+H_{b}+H_{int},\\
				&H_{s}=\varepsilon c^{\dag}c
				+\omega_{0}(b_{0}^{\dag}b_{0}+\frac{1}{2})
				+c^{\dag}c
				\left(
				\omega_{1}(b_{1}^{\dag}b_{1}+\frac{1}{2})
				-\omega_{0}(b_{0}^{\dag}b_{0}+\frac{1}{2})\right),\\
				&H_{b}=\sum_{k}\omega_{k}(a_{k}^{\dag}a_{k}+\frac{1}{2})
			\end{aligned}
		\end{equation}
		where $\omega_{1}=\omega_{0}\sqrt{1+W}$
		and 
		\m{the parameter $W$, satisfying $W\cdot{\bf I}=(\frac{b_{1}^{\dag}+b_{1}}
		{b_{0}^{\dag}+b_{0}})^4-{\bf I}$, is related to the ratio of 
		position quadrature operator
	and	closely related to the generation of squeezed states}\cite{squee}.
	\m{The squeezed bosonic mode $b_{1}^{\dag}$ is defined through  Bogoliubov transformation with the dimensionless quadratic coupling constant $W$,
		\begin{equation} 
			\begin{aligned}
			b_{1}^{\dag}	=\frac{1}{2(1+W)^{1/4}}
				\left((\sqrt{1+W}+1) b_{0}^{\dag}
				+(\sqrt{1+W}-1) b_{0}\right),
			\end{aligned}
		\end{equation}}
		where $b_{0}^{\dag}b_{0}$ and $b_{1}^{\dag}b_{1}$ are the boson number operators when the system is in its ground and excited electronic state, respectively.
		The Boson number operator is
		\begin{equation} 
			\begin{aligned}
				&b_{1}^{\dag}b_{1}
				=\frac{1}{4\sqrt{1+W}}
				\left((4+2W)b_{0}^{\dag}b_{0}
				+W((b_{0}^{\dag})^2+(b_{0})^2)
				+(\sqrt{1+W}-1)^2\right).
			\end{aligned}
		\end{equation}
		Thus the last term in $H_s$ can be rewritten as (see Appendix.A)
		\begin{equation} 
			\begin{aligned}
				c^{\dag}c
				\left(
				\omega_{1}(b_{1}^{\dag}b_{1}+\frac{1}{2})
				-\omega_{0}(b_{0}^{\dag}b_{0}+\frac{1}{2})\right)
				=c^{\dag}c
				\frac{W\omega_{0}}{4}(b_{0}^{\dag}+b_{0})^2,
			\end{aligned}
		\end{equation}
		where the generator of squeeze operator satisfies
		\begin{equation} 
			\begin{aligned}
				&\frac{W}{4}(b_{0}^{\dag}+b_{0})^2
				=\sqrt{1+W}(b^{\dag}_{1}b_{1}+\frac{1}{2})
				-(b^{\dag}_{0}b_{0}+\frac{1}{2}),
			\end{aligned}
		\end{equation}
		
		For system-bath coupling, we consider the electronic-state-dependent coupling
		and electronic-state-independent coupling,
		\begin{equation} 
			\begin{aligned}
				&	H_{int}=H_{int;1}+H_{int;2},\\
				&H_{int;1}=\sum_{k}cc^{\dag}(b_{0}^{\dag}+b_{0}) (a_{k}^{\dag}+a_{k})+
				\sum_{k}c^{\dag}c(b_{1}^{\dag}+b_{1})
				(a_{k}^{\dag}+a_{k}),\\
				&H_{int;2}=\sum_{k}(b_{0}^{\dag}+b_{0}) (a_{k}^{\dag}+a_{k})+
				\sum_{k}(b_{1}^{\dag}+b_{1})
				(a_{k}^{\dag}+a_{k}),\\
			\end{aligned}
		\end{equation}
		where $cc^{\dag}$ and $c^{\dag}c$ play the role of projectors for ground and excited electronic states, respectively.
		$H_{int;1}$ reflects the
		non-adiabatic effect where Born-Oppenheimer approximation breaks down and the nuclear vibrational mode is coupled to the electronic state,
		where the first and second terms couples the ground and excited electronic state to local boson mode and bath, respectively.
		While $H_{int;2}$ involving the Born-Oppenheimer effect where the nuclear vibrational degrees of freedom are treated decoupled from electronic ones,
		and the electrons can instantaneously adjust their state to the changing positions of the much heavier, slower-moving bosonic degree-of-freedom (like the  vibrational motion of the atoms or nuclei).
		
		\section{transitions and the role of complex $W$}
		Now there are four levels:
		$|1\rangle=|0\rangle_{e}|0\rangle_{0}$,
		$|2\rangle=|1\rangle_{e}|0\rangle_{1}$,
		$|3\rangle=|0\rangle_{e}|1\rangle_{0}$,
		$|4\rangle=|1\rangle_{e}|1\rangle_{1}$.
		The transitions involving only the excitation of local boson mode are mediated by $H_{int;1}$ or $H_{int;2}$:
		$|1\rangle \leftrightarrow |3\rangle$,
		$|2\rangle \leftrightarrow |4\rangle$.
		The transitions involving both the excitation of electronic state and local boson mode can only be mediated by $H_{int;2}$:
		$|1\rangle \leftrightarrow |4\rangle$
		(phonon sideband transition),
		$|2\rangle \leftrightarrow |3\rangle$
		(phonon sideband transition).
		The transitions
		$|1\rangle \leftrightarrow |2\rangle$
		(zero-phonon line),
		$|3\rangle \leftrightarrow |4\rangle$
		(phonon sideband transition),
		are induced by the driving light field resonant with the electronic energy gap.
		
		For complex $W$,
		the creational operator of boson in excited electronic state can be written in
		Bogoliubov quasiparticle representation
		(combining the loss and gain operators) as
		\begin{equation} 
			\begin{aligned}
				b^{\dag}_{1}=ub_{0}^{\dag}+vb_{0},
			\end{aligned}
		\end{equation}
		where $u=\frac{\sqrt{1+W}+1}{2(1+W)^{1/4}}$,
		$v=\frac{\sqrt{1+W}-1}{2(1+W)^{1/4}}$,
		and the corresponding boson number operator reads
		\begin{equation} 
			\begin{aligned}
				b^{\dag}_{1}b_{1}=uu^{*}b_{0}^{\dag}b_{0}
				+uv^{*}(b_{0}^{\dag})^2
				+vu^{*}b_{0}^2
				+vv^{*}b_{0}b_{0}^{\dag},
			\end{aligned}
		\end{equation}
		where
		\begin{equation} 
			\begin{aligned}
				&uu^{*}=
				\frac{|1+W|+1+2 {\rm Re}[\sqrt{1+W}]}
				{4\sqrt{|1+W|}},\\
				&uv^{*}=
				\frac{|1+W|-1-2i {\rm Im}[\sqrt{1+W}]}
				{4\sqrt{|1+W|}},\\
				&vu^{*}=
				\frac{|1+W|-1+2i {\rm Im}[\sqrt{1+W}]}
				{4\sqrt{|1+W|}},\\
				&vv^{*}=
				\frac{|1+W|+1-2 {\rm Re}[\sqrt{1+W}]}
				{4\sqrt{|1+W|}},
			\end{aligned}
		\end{equation}
		Note that $|\sqrt{1+W}+1|^2
		=|2+W+2\sqrt{1+W}|
		=|1+W|+2{\rm Re}\sqrt{1+W}+1
		\neq
		2+|W|+2|\sqrt{1+W}|$ and
		$\sqrt{1+W}+(\sqrt{1+W})^*=2{\rm Re}\sqrt{1+W}$,
		$\sqrt{1+W}-(\sqrt{1+W})^*=2i{\rm Im}\sqrt{1+W}$.
		For $\omega_{1}=\omega_{0}\sqrt{|1+W|}$,
		the energy difference between the two local boson modes (squeezed term of $H_{s}$) becomes
		\begin{equation} 
			\begin{aligned}
				\omega_{1}(b_{1}^{\dag}b_{1}+\frac{1}{2})
				-\omega_{0}(b_{0}^{\dag}b_{0}+\frac{1}{2})
				=&\omega_{0}\frac{|1+W|-1}{2}
				b_{0}^{\dag}b_{0}\\
				&
				+\frac{\omega_0 (|1+W|-1+2i{\rm Im}\sqrt{1+W})}{4}b_{0}^2\\
				&+\frac{\omega_0 (|1+W|-1-2i{\rm Im}\sqrt{1+W})}{4}(b_{0}^{\dag})^2\\
				&+\frac{\omega_{0}(|1+W|+1)}{4}
				+\frac{\omega_{0}\sqrt{1+W}}{4}
				-\frac{\omega_{0}(\sqrt{1+W})^*}{4}
				-\frac{\omega_{0}}{2}\\
				=&		\frac{(|1+W|-1)\omega_{0}}{4}(b_{0}^{\dag}+b_{0})^2\\
				&
				+\frac{-i{\rm Im}\sqrt{1+W}\omega_{0}}{2}(b_{0}^{\dag})^2
				+\frac{i{\rm Im}\sqrt{1+W}\omega_{0}}{2}b_{0}^2
				\\
				&
				+\frac{\omega_{0}(|1+W|+1)}{4}
				-\frac{\omega_{0}(|1+W|-1)}{4}\\
				&+\frac{\omega_{0}\sqrt{1+W}}{4}
				-\frac{\omega_{0}(\sqrt{1+W})^*}{4}
				-\frac{\omega_{0}}{2},
				\label{10}
			\end{aligned}
		\end{equation}
		which reduce to $\frac{\omega_{0}W}{4}
		(b_{0}^{\dag}+b_{0})^2$ when $W$ is real.
		
		The terms related to the coordinate and momentum operator read
		\begin{equation} 
			\begin{aligned}
				&b_{1}^{\dag}+b_{1}=(1+W)^{1/4}
				(b_{0}^{\dag}+b_{0}),\\
				&b_{1}^{\dag}-b_{1}=\frac{1}{(1+W)^{1/4}}
				(b_{0}^{\dag}-b_{0}),
			\end{aligned}
		\end{equation}
		Thus in adiabatic approximation, 
		momentum term of boson modes are independent of the electronic state, while the position term is shifted,
		\begin{equation} 
			\begin{aligned}
				&p_{0}=i\sqrt{\frac{m\omega_{0}}{2}}
				(b_{0}^{\dag}-b_{0}),\\
				&p_{1}=i\sqrt{\frac{m\omega_{1}}{2}}
				(b_{1}^{\dag}-b_{1}),\\
				&x_{0}=\sqrt{\frac{1}{2m\omega_{0}}}
				(b_{0}^{\dag}+b_{0}),\\
				&x_{1}=\sqrt{\frac{1}{2m\omega_{1}}}
				(b_{1}^{\dag}+b_{1}),
			\end{aligned}
		\end{equation}
		In adiabatic approximation,
		the jump between distinct potential energy surfaces for the ground and excited states
		is forbidened.
		The nuclear kinetic energy plus the potential energy defined by a single electronic state (ground or excited)
		\begin{equation} 
			\begin{aligned}
				&\frac{p_{0}^2}{2m}+\frac{1}{2}m\omega_{0}^2 x_{0}^2=\omega_{0}(b^{\dag}_{0}b_{0}+\frac{1}{2}),\\
				&\frac{p_{1}^2}{2m}+\frac{1}{2}m\omega_{1}^2 x_{1}^2=\omega_{1}(b^{\dag}_{1}b_{1}+\frac{1}{2}).
			\end{aligned}
		\end{equation}
		The energy difference between the two boson modes in adiabatic approximation is
		\begin{equation} 
			\begin{aligned}
				&\Delta E:=(\frac{p_{1}^2}{2m}+\frac{1}{2}m\omega_{1}^2 x_{1}^2)-(\frac{p_{0}^2}{2m}+\frac{1}{2}m\omega_{0}^2 x_{0}^2)
				\\&	=
				-\frac{\omega_0}{4}(b_{0}^{\dag}-b_{0})^2
				\left(\frac{|1+W|^{1/2}}{(1+W)^{1/2}}-1\right)+
				\frac{1}{2m}\omega_0^2(|1+W|(x_0+d)^2-x_0^2)
				\\&=
				-\frac{\omega_0}{4}(b_{0}^{\dag}-b_{0})^2
				\left(\frac{|1+W|^{1/2}}{(1+W)^{1/2}}-1\right)\\
				&+\frac{\omega_0}{4}(|1+W|-1)
				(b_{0}^{\dag}+b_{0})^2
				+d\omega_{0}^{2}|1+W|\sqrt{\frac{m}{2\omega_{0}}}(b_0^{\dag}+b_0)
				+\frac{1}{2}m\omega_0^2|1+W|d^2.
				\label{14}
			\end{aligned}
		\end{equation}
		Combined with the result of Eq.(\ref{10}),
		we obtain
		\begin{equation} 
			\begin{aligned}
				d=\frac{-c_{1}\pm\sqrt{c_{1}^2-2m\omega_{0}^{2}|1+W|c_{2}}}{m\omega_{0}^{2}|1+W|}
				\label{15}
			\end{aligned}
		\end{equation}
	\m{	where
the oparameter representing the linear displacement coupling acting on the oscillator ($c_1$) and the effective potential correction (including non-Hermitian contributions) that the displacement $d$ must balance ($c_2$) read}
		\begin{equation} 
			\begin{aligned}
				&c_{1}=\omega_0^2|1+W|\sqrt{\frac{m}{2\omega_0}}(b^{\dag}_{0}+b_{0}),\\
				&c_{2}=
				-\frac{\omega_0}{4}(b_{0}^{\dag}-b_{0})^2
				\left(\frac{|1+W|^{1/2}}{(1+W)^{1/2}}-1\right)\\
				&-\left(
				\frac{-i{\rm Im}\sqrt{1+W}\omega_{0}}{2}(b_{0}^{\dag})^2
				+\frac{i{\rm Im}\sqrt{1+W}\omega_{0}}{2}b_{0}^2
				+\frac{\omega_{0}\sqrt{1+W}}{4}
				-\frac{\omega_{0}(\sqrt{1+W})^*}{4}\right).
			\end{aligned}
		\end{equation}
		Then for real $W\ge -1$,
		there are two solutions for
		$dc_{1}+\frac{1}{2}m\omega_0^2|1+W|d^2=0$:
		$d=0$ and $d=-2x_{0}$.
		The solution of $d=0$ is physically possible for the symmetry-preserving transition or the transition to non-bonding state, 
		which is not the case of this article.
		For real $W< -1$,
		Eq.(\ref{15}) is still valid but replacing
		$\sqrt{1+W}=i\sqrt{-(1+W)}$,
		$|1+W|=-(1+W)$.

		Meanwhile, the above boson energy difference Eq.(\ref{14}) is equivalents to
		\begin{equation} 
			\begin{aligned}
				\Delta E=\frac{\omega_1}{2}
				((b_1^{\dag})^2+b_1^2)-\frac{\omega_0}{2}
				((b_0^{\dag})^2+b_0^2),
			\end{aligned}
		\end{equation}
		with
		\begin{equation} 
			\begin{aligned}
				&\frac{p^2_{1}}{2m}=-\frac{\omega_1}{4}(b_{1}^{\dag}-b_{1})^2,
				\frac{p^2_{0}}{2m}=-\frac{\omega_0}{4}(b_{0}^{\dag}-b_{0})^2,\\
				&\frac{1}{2}m\omega_0^2 x_0^2=\frac{\omega_0}{4}(b_0^{\dag}+b_0)^2,
				\frac{1}{2}m\omega_1^2 x_1^2=\frac{\omega_1}{4}(b_1^{\dag}+b_1)^2.
			\end{aligned}
		\end{equation}

		Thus in Born-Oppenheimer (adiabatic) approximation with fast-moving electrons and the slow-moving bosons, where the electronic and vibrational degrees of freedom are separable and the system cannot jump from one electronic state to another,
		$p_{0}=p_{1}$,
		and $x_{1}=x_{0}$ with position shift
		$d=0$.
		This corresponds to the case that the $(1+W)$ is a positive real number.
		As shown in Fig.1,
		according to Franck-Condon principle, since the transition is fast enough that the boson state doesn't have time to change,
		the wavepacket of $b_{0}^{\dag}b_{0}$ is 
		$|\psi\rangle_{0}=|0\rangle_{0}$,
		while the wavepacket of $b_{1}^{\dag}b_{1}$ is 
		$|\psi\rangle_{1}=C_{0}|0\rangle_{1}
		+C_{1}|1\rangle_{1}$ with the Franck-Condon factors $C_{0}=\langle 0|_{1}|0\rangle_{0}$ and $C_{1}=\langle 1|_{1}|0\rangle_{0}$.
		$|0\rangle_{0}$ is the vacuum state with $b_{0}|0\rangle_{0}=0$, $b^{\dag}_{0}|0\rangle_{0}=|1\rangle_0$.
		$|0\rangle_{1}$ is the squeezed vacuum state with $b_{1}|0\rangle_{1}=0$,
		and it is not a minimum uncertainty state,
		with unequal
		position and momentum variances.
		Thus $|0\rangle_{1}$ and $|0\rangle_{0}$ cannot be connected through a displacement operator, instead, $|0\rangle_{1}$ is connected to $|0\rangle_{0}$ through the
		squeezing operation that generates photons in pairs.
		As shown in Fig.1,
		the boson wavepacket $|\psi_{0}\rangle$ instantly projected onto the excited state's potential surface, 
		and becomes a superposition of the new potential's eigenstates, and the coefficients of this superposition are the Franck-Condon factors.
		
		$H_{int;1}$ include the coupled electron-boson coupling term contributes to the off-diagonal effect which
		connecting the ground and excited electronic states, and contributes to the break down of adiabatic approximation
		like the avoided crossings or conical intersections.
		The latter case is related to the complex
		$W$. For complex $W$,
		\begin{equation} 
			\begin{aligned}
				&
				p_{1}=p_{0}\frac{|1+W|^{1/4}}{(1+W)^{1/4}}
				=p_{0}e^{-i\frac{\phi}{4}},\\
				&
				d=x_{1}-x_{0}=
				(\frac{(1+W)^{1/4}}{2|1+W|^{1/4}}-1)
				x_{0}
				=(\frac{1}{2}e^{i\frac{\phi}{4}}-1)
				x_{0}
			\end{aligned}
		\end{equation}
		where $\phi={\rm Arg}(1+W)+2k\pi$.

		\begin{figure}[!ht]
			\centering
			\centering
			\begin{center}
				\includegraphics*[width=0.3\linewidth]{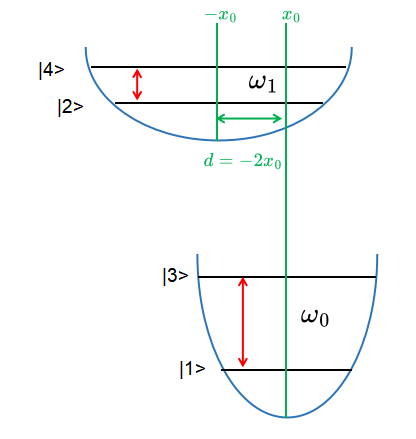}
				\caption{
					The horizontal lines in each electronic potential well represent the quantized  vibrational energy levels.
					The upper well is the excited electronic surface, and the lower well is the ground  electronic surface.
					The two potential energy surfaces have different curvatures:
					The lower well is wider than the upper well
					(and $\omega_{0}>\omega_{1}$) since the bond becomes weaker during the transition from ground electronic state to excited electronic state.
					In ground electronic state, we consider the boson wavepacket locates in its lowest vibrational level at equilibrium position.
					In excited electronic state, we consider the boson wavepacket locates above the lowest vibrational level at nonequilibrium position.
				}
			\end{center}
		\end{figure}

		\section{dissipation}
		\subsection{Hermitian case}
		We next rewrite the $H$ in matrix form whose dimension is $D_{H}=8\times 2^{N}$ with $N$ the number of bath boson modes,
		\begin{equation} 
			\begin{aligned}
				&		H=H_{s}+H_{b}+H_{int},\\
				&H_{s}=\varepsilon c^{\dag}c
				\otimes {\bf I}_{b_{0}}
				\otimes {\bf I}_{b_{1}}
				\otimes_{i=1}^{N}
				{\bf I}_{a_{i}}
				+\omega_{0}(b_{0}^{\dag}b_{0}+\frac{1}{2})
				\otimes {\bf I}_{c}
				\otimes {\bf I}_{b_{1}}
				\otimes_{i=1}^{N}
				{\bf I}_{a_{i}}\\
				&
				+c^{\dag}c\otimes
				\left(
				\omega_{1}(b_{1}^{\dag}b_{1}+\frac{1}{2})
				\otimes {\bf I}_{b_{0}}
				\otimes_{i=1}^{N}
				{\bf I}_{a_{i}}
				-\omega_{0}(b_{0}^{\dag}b_{0}+\frac{1}{2})
				\otimes {\bf I}_{b_{1}}
				\otimes_{i=1}^{N}
				{\bf I}_{a_{i}}\right),\\
				&H_{b}=\sum_{k}\omega_{k}(a_{k}^{\dag}a_{k}+\frac{1}{2})
				\otimes_{i\neq k}
				{\bf I}_{a_{i}}
				\otimes {\bf I}_{c}
				\otimes {\bf I}_{b_{0}}
				\otimes {\bf I}_{b_{1}},\\
				&	H_{int}=H_{int;1}+H_{int;2},\\
				&H_{int;1}=\sum_{k}cc^{\dag}\otimes(b_{0}^{\dag}+b_{0})\otimes (a_{k}^{\dag}+a_{k})	\otimes_{i\neq k}
				{\bf I}_{a_{i}}
				\otimes {\bf I}_{b_{1}}+
				\sum_{k}c^{\dag}c\otimes(b_{1}^{\dag}+b_{1})\otimes
				(a_{k}^{\dag}+a_{k})	\otimes_{i\neq k}
				{\bf I}_{a_{i}}
				\otimes {\bf I}_{b_{0}},\\
				&H_{int;2}=\sum_{k}(b_{0}^{\dag}+b_{0})\otimes (a_{k}^{\dag}+a_{k})	\otimes_{i\neq k}
				{\bf I}_{a_{i}}
				\otimes {\bf I}_{c}
				\otimes {\bf I}_{b_{1}}+
				\sum_{k}(b_{1}^{\dag}+b_{1})\otimes
				(a_{k}^{\dag}+a_{k})	\otimes_{i\neq k}
				{\bf I}_{a_{i}}
				\otimes {\bf I}_{c}
				\otimes {\bf I}_{b_{0}},\\
			\end{aligned}
		\end{equation}
		The Lindblad master equation reads
		\begin{equation} 
			\begin{aligned}
				\frac{d}{dt}\rho=-i[H,\rho]
				+\sum_{i,j=0,1}(
				b_{i}\rho b_{j}^{\dag}-\frac{1}{2}\{b_{j}^{\dag}b_{i},\rho\}),
				\label{21}
			\end{aligned}
		\end{equation}
		where we use
		$b_{0}=c=a_{i}=\frac{1}{2}(\sigma_{x}+i\sigma_{y})$,
		$b_{0}^{\dag}=c^{\dag}=a^{\dag}_{i}=\frac{1}{2}(\sigma_{x}-i\sigma_{y})$
		in truncated Hilbert space.
		$b_{i}\rho b_{j}^{\dag}$ is the quantum jump term,
		which describes the repopulation of states through the transition from a higher-energy state to a lower-energy one, and induce the irreversible flow of probability.
		Using the initial density $\rho_{0}=|\Psi_{0}\rangle\langle\Psi_{0}|$
		with $|\Psi_{0}\rangle$ the ground state of $H$, the time-dependent density can be obtained by solving
		\begin{equation} 
			\begin{aligned}
				\frac{d}{dt}|\rho(t)\rangle\rangle =\mathcal{L}|\rho(t)\rangle\rangle,
			\end{aligned}
		\end{equation}
		where we use the double key notation to denote the vectorized form of matrix, and
		the Liouvillian superoperator reads
		\begin{equation} 
			\begin{aligned}
				\mathcal{L}=-i({\bf I}_{D_{H}}\otimes H-H^{T}\otimes {\bf I}_{D_{H}})
				+\sum_{i,j=0,1}\gamma(b_{i}^{\dag T}\otimes b_{j}
				-\frac{1}{2}{\bf I}_{D_{H}}\otimes b_{i}^{\dag}b_{j}
				-\frac{1}{2}(b_{i}^{\dag}b_{j})^{T} \otimes
				{\bf I}_{D_{H}}).
			\end{aligned}
		\end{equation}

		\begin{figure}[!ht]
			\centering
			\centering
			\begin{center}
				\includegraphics*[width=1\linewidth]{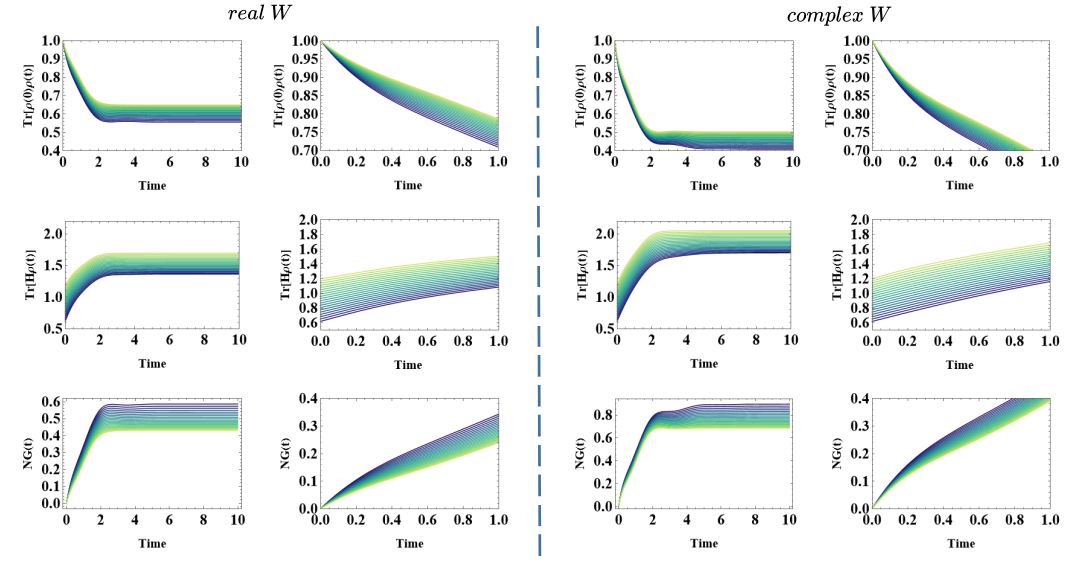}
				\caption{ ${\rm Tr}[\rho(0)\rho(t)]$,
					expectation of $H$, and  rate function $G(t)$ according to the lindblad master equation Eq.(\ref{21}) in Hermitian case.
					We set $\epsilon=1.5$, $\omega_{1}=0.8$,
					$\omega_{k}=1.2$,
					$\gamma=1$.
					The color changing from yellow to blue correspond to the increase of $\omega_{0}$ from 
					$\omega_{0}=\omega_{1}$ to $\omega_{0}=\omega_{1}+1$.
					The left panels correspond to real $W=1.5$, and the right panels correspond to complex 
					$W=1.5+10i$.
					A transition from exponential decay to power law decay can be observed for the survival probability decay rate $G(t)$ in early time.
				}
				\label{bbH}
			\end{center}
		\end{figure}
		\begin{figure}[!ht]
			\centering
			\centering
			\begin{center}
				\includegraphics*[width=1\linewidth]{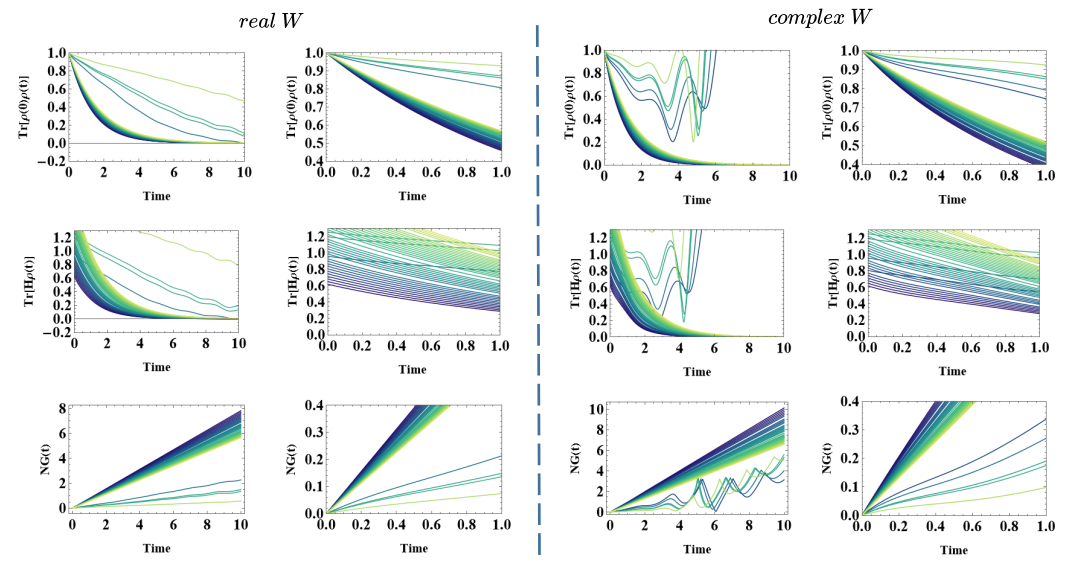}
				\caption{ ${\rm Tr}[\rho(0)\rho(t)]$,
					expectation of $H$, and survival probability decay rate $G(t)$ according to the lindblad master equation about the transitions between the four levels (Eq.(\ref{25})) in Hermitian case.
					A transition from exponential decay to power law decay can be observed for the  rate function $G(t)$ in early time.
					Despite the presence of Liouvilian gap causing the damping effect, the
					sustained oscillations for those specific modes give rise to deviation from the non-exponential decay. Here we set the temperature as $T=\omega_0$.
				}
				\label{transiH}
			\end{center}
		\end{figure}
		In Fig.\ref{bbH}, we choose the ${\rm Tr}[\rho(0)\rho(t)]$ as the probe which is the fidelity between initial state and the time-evolved state. 
	\m{	We further define the survival probability decay rate
		 as
		\begin{equation} 
			\begin{aligned}
				G(t)=-{\rm ln}\frac{{\rm Tr}[\rho(0)\rho(t)]}
				{{\rm Tr}[\rho(0)\rho(0)]}
				=-{\rm ln}{\rm Tr}[\rho(0)\rho(t)].
			\end{aligned}
		\end{equation}
		where ${\rm Tr}[\rho(0)\rho(t)]=\langle \Psi_{0}|\rho(t)|\Psi_{0}\rangle\sim e^{-\Gamma t}$ represents the probability amplitude (or population) of the initial state $|\psi_0\rangle$ remaining in the unnormalized state $\tilde{\rho}(t)$  under the no-jump evolution.
		The time derivative of $G(t)$ is the instantaneous decay rate $\Gamma_{inst}(t) = \frac{dG(t)}{dt} = -\frac{1}{\langle\psi_0| \tilde{\rho}(t) |\psi_0\rangle} \frac{d}{dt}\langle\psi_0| \tilde{\rho}(t) |\psi_0\rangle$.
		If the decay is dominated by a single exponential mode $e^{-\Gamma_{decay} t}$ at long times,  $\frac{dG(t)}{dt}$ will approach a constant value equal to $\Gamma_{decay}$.
	Thus in	linear region, $G(t) \sim \Gamma_{decay} t$, with $\Gamma_{decay}$  the decay rate of the initial state component under no-jump evolution which is related to the imaginary parts of the eigenvalues of $H_{eff}$ or the real parts of the eigenvalues of the corresponding non-Hermitian Liouvillian.
		This is reminiscent with the definition of Loschmidt rate function
		$\lim_{N\rightarrow\infty}
		\frac{-1}{N}{\rm ln}{\rm Tr}[\rho(0)\rho(t)]$, but the latter is for thermodynamic limit.}
		The imaginary part of complex coupling $W$ is closely related to dissipative process with gain or loss.
		As shown in Figs.\ref{bbH},
		the Lindblad equation describes for the transitions between bosonic operators describes a trace-preserving, and stable Markovian process,
		where the Liouvillian has multiple steady states with zero eigenvalue,
		and all the non-zero eigenvalues have negative real part.
	\m{	These multiple steady states and associated long-lived modes emerge discretely and protected by the quantization of symmetry quantum numbers
		(like the fermion number here), corresponding to the non-decaying steady state ($\rho_{ss}$) or true bound states.}
		
		Considering transitions between the four levels, the Lindblad master equation reads
		\begin{equation} 
			\begin{aligned}
				\frac{d}{dt}\rho=&-i[H,\rho]
				+\Gamma_{e;3\rightarrow 1} (
				|1\rangle\langle 3|\rho  |3\rangle\langle 1|-\frac{1}{2}\{|3\rangle\langle 3|,\rho\})
				+\Gamma_{a;1\rightarrow 3} (
				|3\rangle\langle 1|\rho  |1\rangle\langle 3|-\frac{1}{2}\{|1\rangle\langle 1|,\rho\})\\
				&		+\Gamma_{e;4\rightarrow 2} (
				|2\rangle\langle 4|\rho  |4\rangle\langle 2|-\frac{1}{2}\{|4\rangle\langle 4|,\rho\})
				+\Gamma_{a;2\rightarrow 4} (
				|4\rangle\langle 2|\rho  |2\rangle\langle 4|-\frac{1}{2}\{|2\rangle\langle 2|,\rho\}),
				\label{25}
			\end{aligned}
		\end{equation}
		where the emission rates
		$\Gamma_{e;3\rightarrow 1}=\frac{1}{\tau_{0}}$,
		$\Gamma_{e;4\rightarrow 2}=\frac{1}{\tau_{1}}$,
		and the absorption rates
		$\Gamma_{a;1\rightarrow 3}=\frac{1}{\tau_{0}}
		e^{-\omega_{0}/T}$
		$\Gamma_{a;2\rightarrow 4}=\frac{1}{\tau_{1}}
		e^{-\omega_{1}/T}$.
		The relaxation time satisfy
		$\frac{1}{\tau_{1}}=\frac{1}{\tau_{0}}
		(\frac{1}{2}+\frac{\omega_{1}}{2\omega_{0}})^2=\frac{1}{4\tau_{0}}(1+\sqrt{|1+W})^2$.
		The corresponding Liouvillian superoperator
		reads
		\begin{equation} 
			\begin{aligned}
				\mathcal{L}=&
				-i({\bf I}_{D_{H}}\otimes H-H^{T}\otimes
				{\bf I}_{D_{H}})
				+\Gamma_{e;3\rightarrow 1}
				(|3\rangle\langle 1|^T\otimes|1\rangle\langle 3|-\frac{1}{2} {\bf I}\otimes |3\rangle\langle 3|
				-\frac{1}{2} |3\rangle\langle 3|^T\otimes{\bf I} )\\
				& +\Gamma_{a;1\rightarrow 3}
				(|1\rangle\langle 3|^T\otimes|3\rangle\langle 1|-\frac{1}{2} {\bf I}\otimes |1\rangle\langle 1|
				-\frac{1}{2} |1\rangle\langle 1|^T\otimes{\bf I} )\\
				& +\Gamma_{e;4\rightarrow 2}
				(|4\rangle\langle 2|^T\otimes|2\rangle\langle 4|-\frac{1}{2} {\bf I}\otimes |4\rangle\langle 4|
				-\frac{1}{2} |4\rangle\langle 4|^T\otimes{\bf I} )\\
				& +\Gamma_{a;2\rightarrow 4}
				(|2\rangle\langle 4|^T\otimes|4\rangle\langle 2|-\frac{1}{2} {\bf I}\otimes |2\rangle\langle 2|
				-\frac{1}{2} |2\rangle\langle 2|^T\otimes{\bf I} ).
			\end{aligned}
		\end{equation}
		As shown in Fig.\ref{transiH},
		the lindbladian dynamics for the transitions between the four levels has a great dependence on the difference between the vibrational frequencies $\Delta \omega=\omega_0-\omega_1$.
		There are emergent oscillating modes at a series specific value of $\Delta \omega$.
		The oscillations become more dramatic for complex $W$ as shown in right panel of Fig.\ref{transiH}.
		Compared to Fig.\ref{transinH}, 
		it's evident that such oscillation mainly comes from the jumping effect.
	Survival probability decay rate $G(t)$ oscillating in the long-time implies that the system's dynamics are not fully ergodic or thermalizing. Instead, it suggests the presence of quantum revivals and coherent oscillations,
		as well as the dissipative quantum phase transition governed by the finite Liouvillian gap.
		A special phenomenon occurs for the case of transitions between the four levels in Hermitian case,
		where there is diverging and trace-nonpreserving modes,
		which are directly related to the positive real part of the Liouvilian eigenvalues, which violates the detailed balance and complete positivity and trace preservation requirements,
		and implies the instability of the system and the potential net contribution from gain processes.
		The fidelity with initial density increases with time and exhibits oscillations in both the expectation of $H$ and survival probability decay rate. These phenomena  is reminiscent of the adiabatic effect contributed from $H_{int;2}$, even though the term $H_{int;2}$ is not considered here.
		The dependence on $\Delta\omega$ is reduced for higher temperature as shown in Fig.\ref{transiHT}.
		
	\m{	In Fig.\ref{map}, we compare the Liouvillians of the two cases, corresponding to the transitions between local boson operators and between the four levels.
The Liouvillian is block-diagonal only for the former case, with respect to fermion number (eigenvalues of $c^{\dag}c$, 1 or 0) as a strong symmetry which commutes with both the Hamiltonian and jump operators.
In Fig.\ref{map}(a), the two-block structure can be seen, reflecting that the decoupling between the two subspaces.
While in (b), there are coherences between the two subspace.
Further, despite no shown here, we found that the Liouvillians exhibit discrete and relatively continous nature for the transitions between boson operator and the four levels, respectively,
where for latter case the dissipation directly causes transitions and decoherence between these four levels.}

	\m{
	Also, the absence of visible oscillations suggests that the dissipative dynamics (decay) dominate over the coherent dynamics (oscillations). If the decay rates (the real parts of the Liouvillian eigenvalues) are much larger than the oscillation frequencies (imaginary parts of the Liouvillian eigenvalues), the system undergoes overdamped relaxation. Coherent oscillations are quickly suppressed, and the overlap decays smoothly and monotonically towards zero, with the final state orthogonal to the initial state.
}

		\begin{figure}[!ht]
			\centering
			\centering
			\begin{center}
				\includegraphics*[width=1\linewidth]{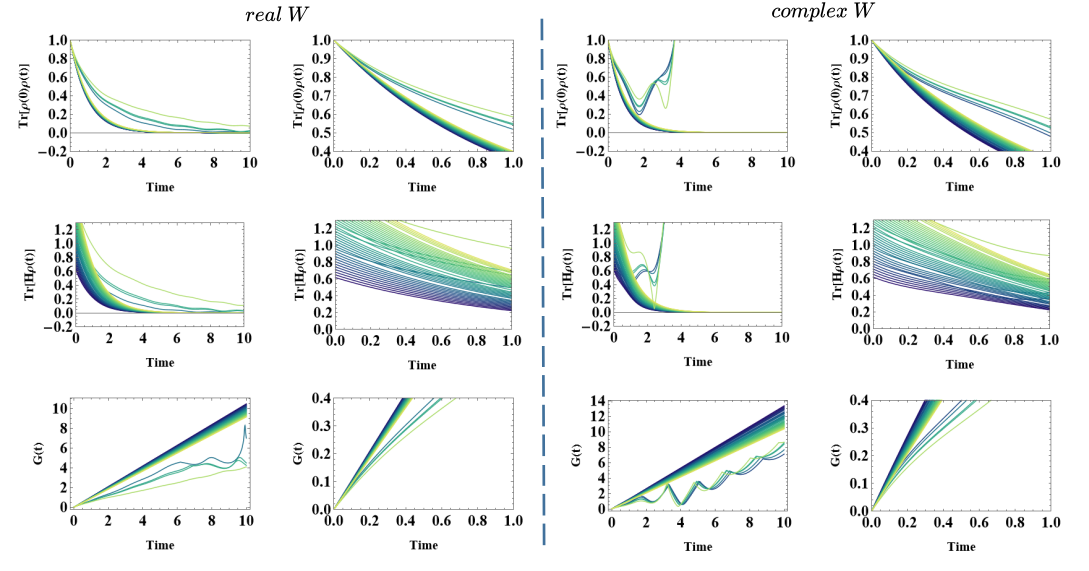}
				\caption{The same with Fig.\ref{transiH} but at higher temperature $T=10\omega_0$.
				}
				\label{transiHT}
			\end{center}
		\end{figure}
		
			\begin{figure}[!ht]
			\centering
			\centering
			\begin{center}
				\includegraphics*[width=1\linewidth]{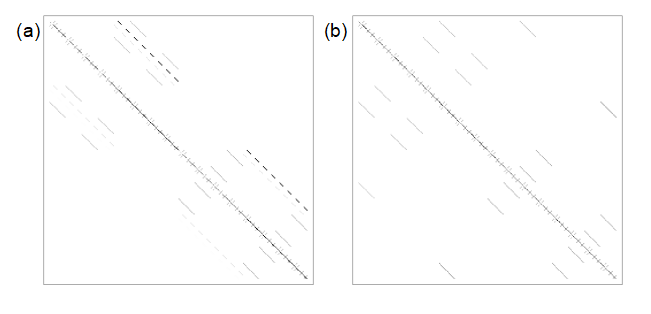}
				\caption{Schematic of Liouvalian in Hermitian case
					for transitions between boson operators (a) and between the four levels (b). 
				}
				\label{map}
			\end{center}
		\end{figure}
		
		\begin{figure}[!ht]
			\centering
			\centering
			\begin{center}
				\includegraphics*[width=1\linewidth]{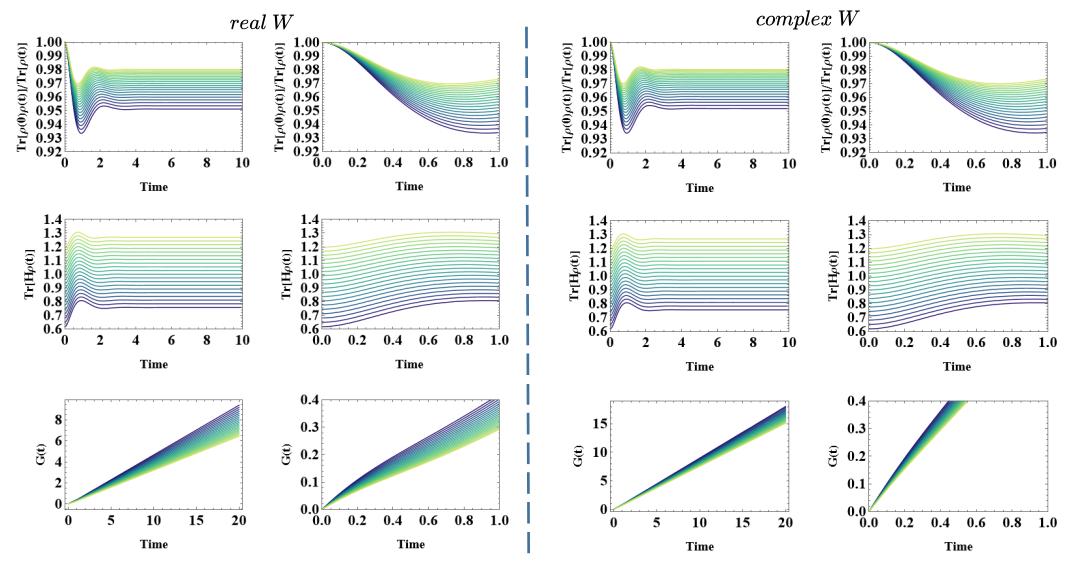}
				\caption{The same with Fig.\ref{bbH} but for non-Hermitian case.
				}
				\label{bbnH}
			\end{center}
		\end{figure}
		
		\begin{figure}[!ht]
			\centering
			\centering
			\begin{center}
				\includegraphics*[width=1\linewidth]{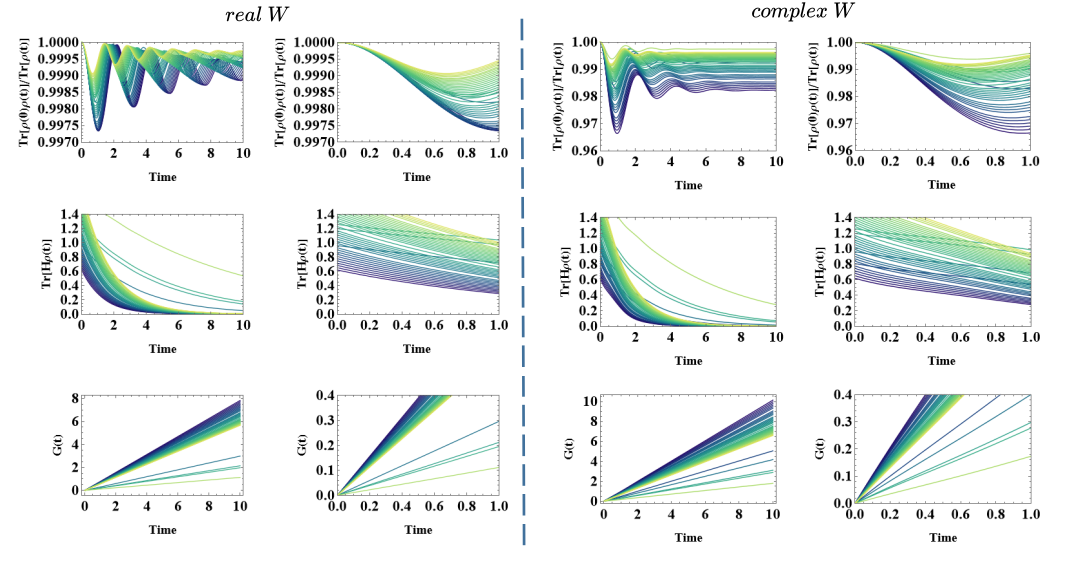}
				\caption{The same with Fig.\ref{transiH} but for non-Hermitian case.
				}
				\label{transinH}
			\end{center}
		\end{figure}
		\begin{figure}[!ht]
			\centering
			\centering
			\begin{center}
				\includegraphics*[width=1\linewidth]{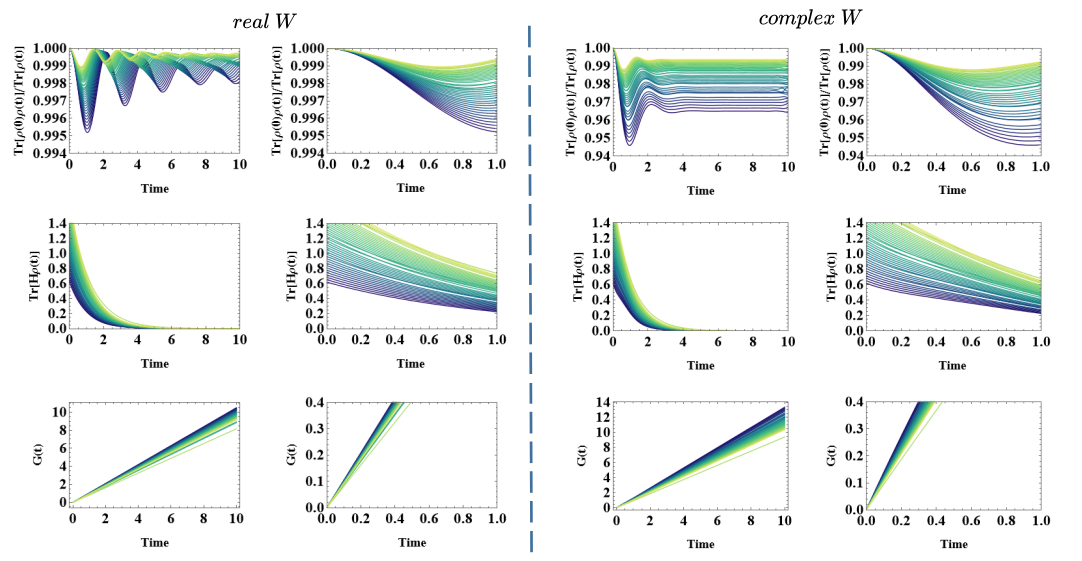}
				\caption{The same with Fig.\ref{transinH} but at higher temperature $T=10\omega_0$.
				}
				\label{transinHT}
			\end{center}
		\end{figure}
		
		\subsection{non-Hermitian case}
		
		When the quantum jump
		processes can be neglected over short timescales,
		the Lindblad master equation Eq.(\ref{21})
		can be rewritten as
		\begin{equation} 
			\begin{aligned}
				\frac{d}{dt}\rho
				=-i(H_{{\rm eff}}\rho-\rho H_{{\rm eff}}^{\dag})
				=-i[H,\rho]-\frac{\gamma}{2}
				\sum_{i,j=0,1}\{b^{\dag}_{i}b_{j},\rho\}
			\end{aligned}
		\end{equation}
		with
		\begin{equation} 
			\begin{aligned}
				H_{{\rm eff}}=H-\frac{i\gamma}{2}
				\sum_{i,j=0,1}b^{\dag}_{i}b_{j}.
			\end{aligned}
		\end{equation}
		Due to the absence of jump term,
		the evolution becomes non-unitary and non-trace-preserving,
		such that $\frac{d}{dt}{\rm Tr}\rho(t)
		=2{\rm Tr}[\rho(t) {\rm Im}H_{{\rm eff}}]=-{\rm Tr}[\rho(t)\sum_{ij}b^{\dag}_{i}b_{j}]$ (note that $\text{Tr}[\rho(t) H_{eff}^\dagger] = \text{Tr}[H_{eff}^\dagger \rho(t)]$).
		Here $b_i^\dagger b_j$ for $i \neq j$ describe dissipative coupling or collective decay processes mediated by the environment
		Specifically, for $i=j$, since $b_i^\dagger b_i$ is positive semi-definite operators (population projectors), $\text{Im}(H_{eff})$ is negative semi-definite, ensuring $\frac{d}{dt} P(t) \le 0$ and thus the decrease of total probability (norm) in the absence of jumps.
		The corresponding Liouvillian superoperator reads
		\begin{equation} 
			\begin{aligned}
				\mathcal{L}
				&=-i({\bf I}_{D_{H}}\otimes H_{{\rm eff}}-(H_{{\rm eff}}^{\dag})^{T}\otimes {\bf I}_{D_{H}})\\
				&=-i({\bf I}_{D_{H}}\otimes H-H^{T}\otimes {\bf I}_{D_{H}})
				-\frac{\gamma}{2}\sum_{i,j=0,1}
				(
				{\bf I}_{D_{H}}\otimes b_{i}^{\dag}b_{j}
				+(b_{i}^{\dag}b_{j})^T\otimes
				{\bf I}_{D_{H}}).
			\end{aligned}
		\end{equation}
		As shown in Figs.\ref{bbnH},\ref{transinH},
		\ref{transinHT},
		when ignore the jump term,
		the effect of boson frequency difference $\omega_{0}-\omega_{1}$ decrease,
		and the survival probability decay rate becomes a linear function of time $G(t)\sim t$
		which means the system's time-dependent fidelity with its initial state (or Loschmidt echo) is decaying exponentially
		$F(t)\sim e^{-ct}$
		and the information loss (ballistic spreading) occurs at a constant rate,
		reflects a process of decoherence and irreversible information loss due to the sensitivity to the perturbations.
		\m{
In linear regime, the exponential decay $F(t):={\rm Tr}[\rho(0)\rho(t)] \sim e^{-\Gamma t}$  is the standard behavior for Markovian open quantum systems with constant decay rate. It implies that the probability of the system not having undergone a jump decreases exponentially over time, governed by a well-defined decay rate (related to the Liouvillian gap or effective Hamiltonian eigenvalues). This is the expected behavior far from critical points or localization effects.
As shown in Fig.\ref{transinHT}, the dependence on $\Delta\omega$ is again reduced at higher temperature and the number of special mode is also reduced.}

		We also notice that,
		the survival probability decay rate initially exhibits power-law behavior $G(t)\sim t^{\alpha}$ with exponent $0<\alpha<1$,
		i.e., power-law decay (polynomial decay) of the overlap $F(t)\sim e^{-ct^{\alpha}}$
		which is slower than exponential decay but faster than a pure polynomial decay,
		as well as the sub-ballistic information spreading.
		\m{This slower-than-exponential decay (sub-exponential) corresponds to the case where transport and relaxation are hindered and the nonMarkovian environments with long memory times can lead to power-law or sub-exponential decay dynamics.
		}
		This also implies a strong coupling to a noisy or dissipative environment (bath with strong long-term memory effect) and exhibits the non-unitary and quantum Zeno effect,
		where the environment's influence effectively performs continuous measurements that project the system back onto its initial state and thus slow down the thermalization.
	The quantum Zeno effect here	persisting past the initial short-time regime and slowing down the decay through frequent or continuous observation.	
		The survival probability decay rate in both the Hermitian and non-Hermitian cases initially follows a power-law behavior with exponent $0<\alpha<1$ and subsequently becomes linear with time, but saturation at long times occurs only in the Hermitian case.
		In all cases, the $G(t)\sim t^{\alpha}$
		for $\alpha>1$ cannot be found,
		thus the Gaussian fidelity decay 
		$F(t)\sim e^{-ct^{\alpha}}$ for $\alpha >1$ is absent, and this case can only be found for the case of
		initial dephasing of wave function due to the spread of different energy components after a quench.
		Note that the above subballistic regime still corresponds to finite Liouvillian gap,
		a zero Liouvillian gap corresponds to the case of $G(t)\sim {\rm ln}t$
		and $F(t)\sim t^{-\alpha}$ with $\alpha>0$,
		in which case there is continuous degrees of freedom or diffusion-like dynamics, such as the many-body localization due to the 
		non-ergodicity.

\m{	For the conditional survival probability $\frac{{\rm Tr}[\rho(0)\rho(t)]}{{\rm Tr}[\rho(t)]}$, which represent the overlap between the initial state and the conditional state, the evolution of the conditional state is primarily driven by the coherent part ($H$) within the effective Hamiltonian $H_{eff}$. This coherent part causes the quantum state to oscillate or rotate in Hilbert space at frequencies related to the energy differences in $H$.
	The conditional survival probability is close to 1 upto longer time, indicates that the conditional state maintains a significant resemblance to the initial state $\rho(0)$. This happens if the decay rates ($\gamma_k$) are relatively small compared to the oscillation frequencies, or if the initial state projects strongly onto eigenmodes of $H_{eff}$ that decay very slowly. The normalization prevents the overlap from simply decaying due to the loss of overall probability $p_0(t) = \text{Tr}[\rho(t)]$,
	and making the decaying amplitude solely originates from non-Hermitian part $-\frac{i}{2}\sum L_k^\dagger L_k$
	(the rate of probability loss during no-jump evolution).
	Figs.\ref{transinH},
	\ref{transinHT} show that
	the oscillation is significantly suppressed for complex $W$.
	This implies that 
	a complex $W$ introduces additional non-Hermiticity into the system's coherent part $H$ (through a squeezing transformation), modifying how decay manifests in the conditional overlap, often enhancing the damping effect on oscillations.
	The effect of complex $W$ on the damping of  oscillations within the conditional state
	can be studied through the non-Hermitian terms $-\frac{i}{2}\sum_{i=1,2,3,4} \gamma_{i}|i\rangle\langle i|$,
	which reads
	\begin{equation} 
		\begin{aligned}
			-\frac{i}{2}\sum_{i=1,2,3,4} \gamma_{i}|i\rangle\langle i| =& -\frac{i}{2\tau_0} [ e^{-\omega_0/T} |1\rangle\langle 1| + \frac{(1+\sqrt{|1+W|})^2}{4} e^{-\omega_0\sqrt{|1+W|}/T} |2\rangle\langle 2|\\
			&  + |3\rangle\langle 3| + \frac{(1+\sqrt{|1+W|})^2}{4} |4\rangle\langle 4| ].
	\end{aligned}			\end{equation} 
	Thus the complex component of $W$ affecting the decay rates $\Gamma_{e;4\to 2}$ and $\Gamma_{a;2\to 4}$.
	Also, the non-unitary Bogoliubov transformation with complex $W$ implies that the relationship between $b_0$ and $b_1$ is influenced by interactions involving gain, loss, or phase effects.
	The anti-commutator term $-\frac{\gamma_{i}}{2}\{|i\rangle\langle i|, \rho\}$ removes probability from the system and causes populations to decay,
	and leads to a decrease in trace $\frac{d}{dt}\text{Tr}[\rho(t)]=\frac{d}{dt}\text{Tr}[\rho(t)]_{decay} = -\sum_i \gamma_i \text{Tr}[|i\rangle\langle i| \rho] < 0$, and it
	causes the nonunitary evolution
	and nonunitary $e^{\mathcal{L}t}$ where $\mathcal{L}$ is not anti-Hermitian.
	While the full Lindblad equation contains the jump term $\frac{d}{dt}\text{Tr}[\rho(t)]_{jump}
	=\text{Tr}\sum_{ji} \gamma_{ji} |j\rangle\langle i|\rho |i\rangle\langle j|= \sum_i \gamma_i \text{Tr}[|i\rangle\langle i| \rho] > 0$,
	where we use the property $\text{Tr}[L_j \rho L_j^\dagger] = \text{Tr}[L_j^\dagger (L_j \rho)] = \text{Tr}[(L_j^\dagger L_j) \rho]=\text{Tr}[ \rho(L_j^\dagger L_j)]$.
}

		A small Liouvillian gap would not produce a true power-law decay in the long time limit; instead, the behavior would eventually transition to the standard exponential decay dictated by the Liouvillian gap.

		The non-Hermitian master equation describes the incoherent and non-unitary evolution due to the absence of jump term.
		Using the relations 
		$|A\rho B\rangle\rangle=(B^{T}\otimes A)|\rho\rangle\rangle$,
		$e^{A\otimes I+I\otimes B}=e^{A}\otimes e^{B}$,
		and $\mathcal{L}=-i({\bf I}_{D_{H}}\otimes H_{{\rm eff}}-(H_{{\rm eff}}^{\dag})^{T}\otimes {\bf I}_{D_{H}})$,
		we have
		\begin{equation} 
			\begin{aligned}
				e^{\mathcal{L}t}=(e^{-i{\bf I}_{D_{H}}\otimes H_{{\rm eff}}t}\otimes {\bf I}_{D_{H}})
				(e^{-i{\bf I}_{D_{H}}\otimes H_{{\rm eff}}t}\otimes {\bf I}_{D_{H}}),
			\end{aligned}
		\end{equation}
		thus $\rho(t)=e^{\mathcal{L}t}\rho(0)
		=e^{-iH_{{\rm eff}}t}\rho(0)e^{iH_{{\rm eff}}^{\dag}t}$.
		The Liouvillian gap $\Delta_{L}={\rm min}_{{\rm Re}\lambda_{i}\neq 0}|{\rm Re}\lambda_{i}|$,
		which determines the lowest decay rate, is also noticeable here.
		Distinct with the Hermitian case
		where the Liouvillian gap is always zero,
		the Liouvillian gap is nonzero for both the 
		transitions between bosonic operators (Eq.\ref{21}) and the transitions between the four levels (Eq.\ref{25}),
		and the Liouvillian gap is related to the imaginary part of $W$, and independent of the temperature $T$.
		
		As shown in Fig.\ref{gapbb},
		in Hermitian case,
		the Liouvillian gap for the transitions between bosonic operators as a function of ${\rm Im}W$ and $\Delta\omega$, is inversely proportional to $\Delta\omega$ and
		reaches its minimal (maximal) value at negative (positive) ${\rm Im}W$.
	\m{This implies the critical slowing down of dissipation and long-lived coherences
		for large $\Delta\omega$.
			Since $\rho(t)$ converges to $\rho_{ss}$ at a rate $\sim e^{-\Delta_{L}t}$, this convergence becomes slower for small $\Delta\omega$.}
		In non-Hermitian case,
		the Liouvillian gap for the transitions between bosonic operators 
		depends only on ${\rm Im}W$,
		which is minimal at ${\rm Im}W=0$,
		and independent of $\Delta\omega$.
\m{In both cases, a non-zero Liouvillian gap always implies better stability of the steady state in the sense that any perturbation will decay exponentially quickly back to the steady state (or zero, in the non-Hermitian case).
}
		
	As shown in Fig.\ref{gaptransi}, for transitions between the four levels, $\Delta_{L}$ depends on the ${\rm Im}W$, $\Delta\omega$ and temperature,
		where the minimal of 
		$\Delta_{L}$ deviate away from ${\rm Im}W=0$
		in Hermitian case
		and locates in ${\rm Im}W=0$
		in non-Hermitian case.
		Thus for the transitions between bosonic operators in non-Hermitian case, we have
		$\Delta_{L}\propto |{\rm Im}W|$,
		which implies
		stronger the system couple to the environment (stronger damping), the faster it loses its initial non-equilibrium state and settles into its long-term steady state,
		and this also apply to the transitions between the four levels as long as $|{\rm Im}W|\rightarrow 0$.
		
		A larger Liouvillian gap corresponds to faster decay of fidelity and Loschmidt rate function increase more rapidly over time.
		A smaller Liouvillian gap corresponds to slower decay of fidelity, reflecting the dissipative phase transition or a systems with conserved quantity, and the survival probability decay rate will increase more slow  (sub-ballistic decay).

		Through the equation $\frac{d\rho(t)}{dt}=\mathcal{L}\rho(t)$,
		we have $|\rho(t)\rangle\rangle
		=e^{\mathcal{L}t}|\rho_{i}\rangle\rangle
		=\sum_{i}c_{i}e^{\lambda_{i}t}|\rho_{i}\rangle\rangle$,
		where $|\rho_{i}\rangle\rangle$ are the eigenvector of $\mathcal{L}$ corresponding to $i$-th eigenvalue, and
		$\rho_{i}$ is the density matrix in Liouvillian formalism whose vectorized form
		is the eigenstate of Liouvilian.
		The coefficients $c_{i}$
		are the overlap between initial state and the left Liouvillian eigenstates $c_{i}=\langle\langle\tilde{\rho}_{i}|\rho(0)\rangle\rangle$,
		with $|\rho(0)\rangle\rangle=\sum_{i}c_{i}|\rho_{i}\rangle\rangle$.
		Since all components $|\rho_{i}\rangle\rangle$
		for $i\neq 0$ decay exponentially, 
		only the steady-state component survives
		in long time limit,
		such that the final steady-state is 
		$|\rho(t\rightarrow\infty)\rangle\rangle
		=c_{0}|\rho_{0}\rangle\rangle
		=|\rho_{ss}\rangle\rangle$.

		For the case that
		the final steady state is orthogonal to the initial state with zero fidelity, the long-time dynamics of $G(t)$ will fully governed by the Liouvillian gap and has no steady-state contributions.
		The Liouvillian superoperators $\mathcal{L}$ are non-Hermitian, 
		their right and left eigenvectors form a biorthogonal basis,
		such that
		$\mathcal{L}|\rho_{i}\rangle\rangle=
		\lambda_{i}|\rho_{i}\rangle\rangle$,
		$\langle\langle\rho_{i}|\mathcal{L}=
		\lambda_{i}\langle\langle\rho_{i}|$,
		$\mathcal{L}^{T}|\tilde{\rho}_{i}\rangle\rangle=
		\lambda_{i}|\tilde{\rho}_{i}\rangle\rangle$,
		with $\langle\langle\tilde{\rho}_{i}|\rho_{j}\rangle\rangle=\delta_{ij}$.
		Since the Liouvillian is diagonalizable (without any Jordan blocks with dimension larger than one),
		its eigenvectors form a complete biorthogonal set,
		and the
		eigenstates (multiple steady states) in a zero-eigenvalue subspace are orthogonal to each other (without the exceptional points).
			\m{Distinct to the persistent oscillation in conditional survival probability $\frac{{\rm Tr}[\rho(0)\rho(t)]}{{\rm Tr}[\rho(t)]}$ in non-Hermitian case,
			for Hermitian case,
			despite the density matrix elements themselves might exhibit damped oscillations, their projection onto the initial state might decay monotonically.
			This can happen if the initial state projection primarily captures the decaying part of the dynamics, while the oscillatory parts live in orthogonal components.
			The projection onto the initial state $\rho(0) = |\Psi_0\rangle\langle\Psi_0|$ is
			\begin{equation} 
				\begin{aligned}
					\text{Tr}[\rho(0)\rho(t)] = \langle\Psi_0|\rho(t)|\Psi_0\rangle
					= \langle\langle \rho(0) | \rho(t) \rangle\rangle = \langle\langle \rho(0) | \sum_i c_i e^{\lambda_i t} |\rho_i\rangle\rangle
					=\sum_i c_i \langle\langle \rho(0) | \rho_i \rangle\rangle e^{\lambda_i t}
			\end{aligned}			\end{equation}
			where we use the Hilbert-Schmidt inner product $\langle\langle A | B \rangle\rangle = {\rm Tr}[A^\dagger B]$  in Liouville space with vectorized operators. The monotonic decay can be explained by the dominating overlap between $|\rho(0)\rangle\rangle$ and $|\rho_n\rangle\rangle$ with ${\rm Im}\lambda_{n}=0$.}

		In both the Hermitian and non-Hermitian cases,
		there are special modes marked by a series of discrete values of $\Delta \omega$.
		These discrete critical values of the detuning $\Delta \omega$ are related to the resonant frequencies of the underlying Hamiltonian $H$,
		i.e., the unitary resonance in Hermitian case where maximal sustained oscillations (coherence) occur for degeneracy in the purely imaginary eigenvalues, 
		and dissipative resonance (dynamical quantum phase transition) for non-Hermitian case.
		The oscillating mode only found in Hermitian case, which is consistent with the finding that,
		the purely imaginary eigenvalue of Liouvilian can only be found in Hermitian case,
		otherwise the oscillations $e^{i{\rm Im}\lambda t}$ damped exponentially with $e^{{\rm Re}\lambda t}$.
		Such resonance or degeneracy in the spectrum of the Liouvillian,
		is suppressed by the increased temperature,
		as shown in Figs.\ref{gaptransi},\ref{transiHT},\ref{transinHT}.
		For both the Hermitian and non-Hermitian cases, the coherent part of Liouvillian $-i[H,\cdot]$ is anti-Hermitian,
		but the full Liouvillian is not anti-Hermitian, and consequently,
		$\mathcal{L}^{\dag}\neq -\mathcal{L}$,
		${\rm Re}\lambda<0$ and
		$e^{\mathcal{L}t}$ is nonunitary.

		We also note the two specific cases:
		First one is the pure loss where the diagonalizable non-Hermitian Liouvillian  has an entirely real spectrum and becomes pseudo-Hermitian with respect to a positive-definite metric $G$, $\mathcal{L}=G^{-1}\mathcal{L}^{\dag}G$.
		Second one is that the dissipation terms have canceled out any loss or gain,
		resulting in zero net dissipation or decoherence.
		Then both the coherence part and dissipation part of the Liouvillian are anti-Hermitian, which leads to:
		$\mathcal{L}$ is anti-Hermitian,
		${\rm Re}\lambda=0$,
		$e^{\mathcal{L}t}$ is unitary.

		\begin{figure}[!ht]
			\centering
			\centering
			\begin{center}
				\includegraphics*[width=0.7\linewidth]{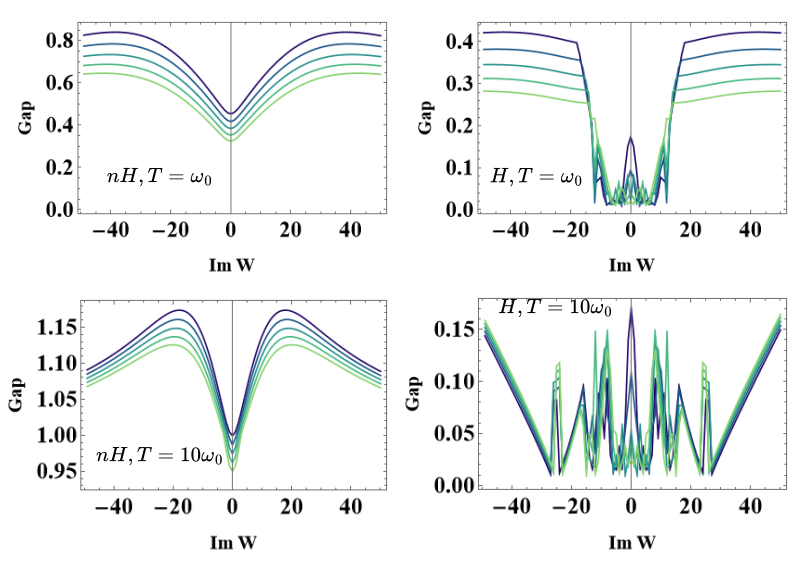}
				\caption{Liouvillian gap for the transitions between four levels
					for Hermitian case ($H$) and non-Hermitian case ($nH$) at different temperature.
					Thus the nullspace with non-defective degeneracy can only be found in the Hermitian case. While for non-Hermitian case, despite the absence of zero eigenvalue, the non-defective degeneracy can be evidenced by the equal algebraic multiplicity geometric multiplicity for certain states (the linear independent eigenstates with coalesced eigenvalues).
				}
				\label{gaptransi}
			\end{center}
		\end{figure}
		\begin{figure}[!ht]
			\centering
			\centering
			\begin{center}
				\includegraphics*[width=0.4\linewidth]{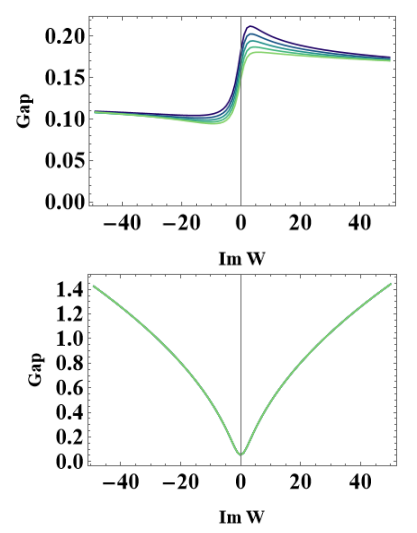}
				\caption{Liouvillian gap for the transitions between bosonic operators in Hermitian (upper panel) non-Hermitian case (lower panel).
				}
				\label{gapbb}
			\end{center}
		\end{figure}

		\subsection{Zero-Phonon Line (ZPL) and Phonon Sideband Transition (PSB)}
		
	\m{	Zero-phonon line (ZPL) and phonon sideband transition (PSB) are available in the presence of driving term $H_{\mathrm{drive}}(t) = -\vec{\mu} \cdot \vec{E}(t) \approx  \Omega_L (c^\dagger e^{-i\omega_L t} + c e^{i\omega_L t}) $ where $\Omega_L$ is the Rabi frequency related to the laser field strength, $\omega_L$ is the laser frequency.}
		
		For ZPL transition $|1\rangle\leftrightarrow|2\rangle$,
		the detuning is $\Delta_{ZPL}=\omega_e-\omega_L$
		with laser frequency $\omega_L$, and
		the Rabi frequency is the product of the laser's driving strength $\Omega_L$ and the Franck-Condon factor 
		$\Omega_{12}=\Omega_{L}
		|\langle 0|_{0}|0\rangle_{1}|^2$.
		Thus the transition rate (effective Rabi frequency) at resonance is
		$\Gamma_{1\leftrightarrow 2}=\Omega_{12}$.
		For PSB trasnition
		$|3\rangle\leftrightarrow|4\rangle$,
		the detuning ar resonance is $\Delta\omega$,
		and the Rabi frequency and transition rate are $\Omega_{34}=\Omega_{L}
		|\langle 1|_{0}|1\rangle_{1}|^2$
		and $\Gamma_{3\leftrightarrow 4}=\frac{\Omega_{34}}{\sqrt{1+\frac{\Delta\omega}{\Omega_{34}}}}$, respectively.
		The Lindblad master equation reads
		\begin{equation} 
			\begin{aligned}
				\frac{d}{dt}\rho=&-i[H,\rho]
				+\Gamma_{1\rightarrow 2} (
				|1\rangle\langle 2|\rho  |2\rangle\langle 1|-\frac{1}{2}\{|2\rangle\langle 2|,\rho\})
				+\Gamma_{1\rightarrow 2} (
				|2\rangle\langle 1|\rho  |1\rangle\langle 2|-\frac{1}{2}\{|1\rangle\langle 1|,\rho\})\\
				&		+\Gamma_{3\rightarrow 4} (
				|3\rangle\langle 4|\rho  |4\rangle\langle 3|-\frac{1}{2}\{|4\rangle\langle 4|,\rho\})
				+\Gamma_{3\rightarrow 4} (
				|4\rangle\langle 3|\rho  |3\rangle\langle 4|-\frac{1}{2}\{|3\rangle\langle 3|,\rho\}),
			\end{aligned}
		\end{equation}
		The corresponding Liouvillian superoperator
		reads
		\begin{equation} 
			\begin{aligned}
				\mathcal{L}=&
				-i({\bf I}_{D_{H}}\otimes H-H^{T}\otimes
				{\bf I}_{D_{H}})
				+\Gamma_{1\rightarrow 2}
				(|1\rangle\langle 2|^T\otimes|2\rangle\langle 1|-\frac{1}{2} {\bf I}\otimes |1\rangle\langle 1|
				-\frac{1}{2} |1\rangle\langle 1|^T\otimes{\bf I} )\\
				& +\Gamma_{1\rightarrow 2}
				(|2\rangle\langle 1|^T\otimes|1\rangle\langle 2|-\frac{1}{2} {\bf I}\otimes |2\rangle\langle 2|
				-\frac{1}{2} |2\rangle\langle 2|^T\otimes{\bf I} )\\
				& +\Gamma_{3\rightarrow 4}
				(|3\rangle\langle 4|^T\otimes|4\rangle\langle 3|-\frac{1}{2} {\bf I}\otimes |3\rangle\langle 3|
				-\frac{1}{2} |3\rangle\langle 3|^T\otimes{\bf I} )\\
				& +\Gamma_{3\rightarrow 4}
				(|4\rangle\langle 3|^T\otimes|3\rangle\langle 4|-\frac{1}{2} {\bf I}\otimes |4\rangle\langle 4|
				-\frac{1}{2} |4\rangle\langle 4|^T\otimes{\bf I} ).
			\end{aligned}
		\end{equation}
		As shown in Figs.\ref{ZPL3},\ref{ZPLgap}, for transitions induced by the driving light field,
		the fidelity, expectation of $H$,
		 survival probability decay rate, and Liouvilian gap are independent of ${\rm Im}W$.
		The Liouvilian gap inversely proportional to $\Delta\omega$ in non-Hermitian case.
		Specifically, the survival probability decay rate initially exhibits $G(t)\sim t^{\alpha}$ 
		with $\alpha>1$ in Hermitian case,
		which corresponds to the super-linear regime characterized by an effectively increasing decay rate, i.e., 
		super-exponential or Gaussian decay at short time scale.
		This is also the regime of 
		anti-Zeno effect where the decay is faster than exponential, and the  environment accelerates decoherence.

		For adiabatic PSB transitions $|1\rangle\leftrightarrow|4\rangle$ and $|2\rangle\leftrightarrow|3\rangle$ contributed by $H_{int;2}$.
The modes characterized by $\Delta\omega$ can be used to distinguish various symmetries and oscillation patterns. 
The survival probability decay rate exhibits $G(t)\sim t^{\alpha}$ 
with $0<\alpha<1$ at short times and either saturates (for the Hermitian case) or oscillates and increases (for the non-Hermitian case) at long times.
Each specific pattern corresponds to
a strong symmetry of the Lindblad dynamics generated by a unitary operator that commutes with the Hamiltonian and jump operator $[\hat{U},H]=[\hat{U},L]=0$, and the corresponding symmetry transformation commutes with the Liouvillian
$[U(\rho),\mathcal{L}(\rho)]=[\hat{U}\rho\hat{U}^{\dag},\mathcal{L}(\rho)]=0$.
Thus $\mathcal{L}$ preserves the eigenvectors of $U$, allowing the Hilbert space of operators (Liouville space) to be decomposed into orthogonal subspaces (invariant under $\mathcal{L}$).
Such operators identify subspaces
in the operator Hilbert space		
that are invariant under
the Lindblad dynamics.
Since the dynamics are decoupled across these invariant subspaces, the observables will be dominated by the Liouvillian eigenvalues within the specific subspace relevant to initial state. 
		
				\begin{figure}[!ht]
			\centering
			\centering
			\begin{center}
				\includegraphics*[width=1.1\linewidth]{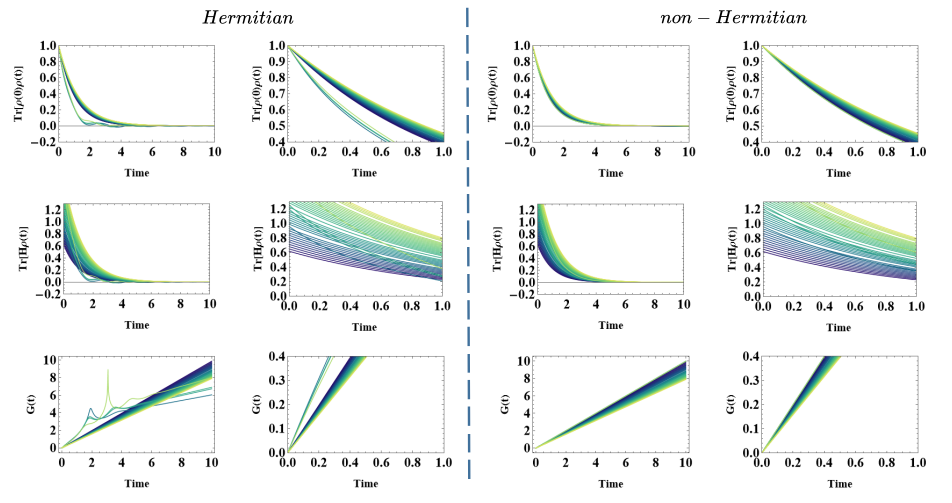}
				\caption{ Hilbert–Schmidt overlap, expectation of $H$, and
					survival probability decay rate, for transitions induced by driving laser in Hermitian case and non-Hermitian case.
				}
				\label{ZPL3}
			\end{center}
		\end{figure}
		\begin{figure}[!ht]
			\centering
			\centering
			\begin{center}
				\includegraphics*[width=0.5\linewidth]{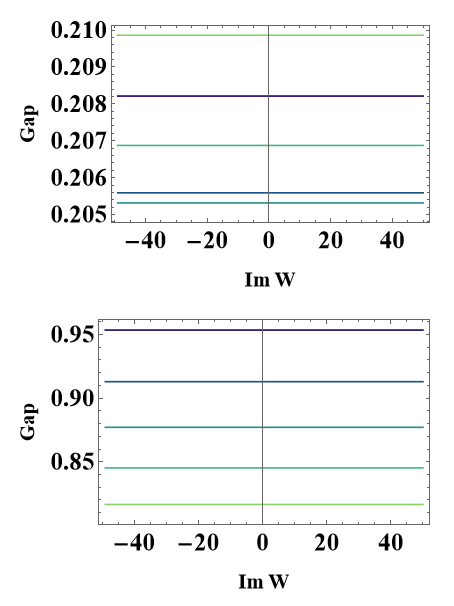}
				\caption{Liouvilian gap for transitions induced by driving laser in Hermitian case (upper panel) and non-Hermitian case (lower panel).
				}
				\label{ZPLgap}
			\end{center}
		\end{figure}
		
		For ensemble average,
		$\rho(t)=\frac{1}{M}\sum_{m}|\psi_{m}(t)\rangle\langle\psi_{m}(t)|$ with $m$ the indices of the $M$ trajectories. $|\psi_{m}(0)=|\Psi_{0}\rangle$ the ground state of $H$.
		$F(t)=\frac{1}{M}\sum_{m}|\langle \Psi_{0}|\psi_{m}(t)\rangle|^2$.
		At short time before jump,
		$|\psi_{m}(\delta t)\rangle=(1-iH_{{\rm eff}}\delta t)|\Psi_{0}\rangle$,
		where the nonunitary evolution
		$\frac{d|\psi_{m}(\delta t)\rangle}{dt}=-iH_{{\rm eff}}|\psi_{m}(\delta t)\rangle$
decreases the norm $\langle\psi_{m}(\delta t)|\psi_{m}(\delta t)\rangle<\langle\Psi_{0}|\Psi_{0}\rangle$.
While the probability of a jump
is $\langle\psi_{m}(t)|\psi_{m}(t)\rangle-\langle\psi_{m}(t+\delta t)|\psi_{m}(t+\delta t)\rangle
=
-i\delta t \langle\psi_{m}(t)|H_{{\rm eff}}^{\dag}-H_{{\rm eff}}|\psi_{m}(t)\rangle
=\delta t\sum_{ij}\langle\psi_{m}(t)|b^{\dag}_{i}b_{j}|\psi_{m}(t)\rangle$.

According to fluctuation-dissipation theorem\cite{Schaller,Shiraishi,Pausch},
the bath correlation in the thermalized reservoir
where $\langle a_{k}^{\dag}a_{k}\rangle
={\rm Tr}[a_{k}^{\dag}a_{k}\rho(0)]
={\rm Tr}_{s}[\rho_{s}(0)]
{\rm Tr}_{b}[a_{k}^{\dag}a_{k}\rho_{b}(0)]
=(e^{\omega_{k}/T}-1)^{-1}$
and ${\rm Tr}[a_{k}^{\dag}a_{k}\rho(t)]
=\frac{1}{M}\lim_{t\rightarrow\infty}\sum_{m}
\langle\psi_{m}(t)|a_{k}^{\dag}a_{k}|\psi_{m}(t)\rangle$,
reads
\begin{equation}
			\begin{aligned}
&\mathcal{C}_{11}(\tau)=\int g(\omega)[(\langle a^{\dag}a\rangle+1)e^{-i\omega\tau}+\langle a^{\dag}a\rangle e^{i\omega\tau}]d\omega
=\int g(\omega)(\frac{e^{-i\omega\tau}}{1-e^{-\omega/T}}+
\frac{e^{i\omega\tau}}{e^{\omega/T}-1})d\omega=\mathcal{C}_{22}(\tau),\\
&\mathcal{C}_{12}(\tau)=-i\ {\rm sgn}(\omega) \mathcal{C}_{11}(\tau)=-\mathcal{C}_{21}(\tau),
			\end{aligned}
\end{equation}
where the spetral density $g(\omega)=\sum_{k}|h_{k}|^2\delta(\omega-\omega_k)\approx|h(\omega)|^2 \mathcal{D}(\omega)
=\frac{dN}{d\omega}|h(\omega)|^2$ with coupling strength $|h_k|^2$ between the system and the $k$-th bath oscillator mode
and we apply $\sum_{k}|h_k|^2 f(\omega_k)
=\int^{\infty}_{0}g(\omega)f(\omega)d\omega$ for a large and continuous bath (thermodynamic limit where $N\rightarrow\infty$).
The small $N$ setted in this article cause the quasi-periodic oscillations (recurrence) instead of irreversible decay.

\subsection{Non-Markovian}
In interacting picture,
$a_{k}(t)=e^{iH_{b}t}a_{k}e^{-iH_{b}t}
=e^{i\omega_{k}t}a_{k}$.
The system-bath linear coupling
for multiple bath mode ($N>1$) is
$h_{int}=h_{s}\otimes A$,
with $h_{s}=c^{\dag}c\otimes(b_{0}^{\dag}+b_{0})+c^{\dag}c\otimes(b_{1}^{\dag}+b_{1})$ the system operator that participate the coupling and $A=\sum_{k}(h_{k}a_{k}^{\dag}+h_{k}^{*}a_{k})$the Hermitian bath operator.
$h_{s}(\omega)=\sum_{E_{m}-E_{n}=\omega}|m\rangle\langle m|h_{s}|n\rangle\langle n|$
where the Bohr frequency $\omega$ is determined by:
$c^{\dag}c\otimes b_{0}^{\dag}
|n_{c},n_{0},n_{1}\rangle=n_{c}\sqrt{n_{0}+1}|n_{c},n_{0}+1,n_{1}\rangle$,
$\omega_{0}=E(n_{c},n_{0}+1,n_{1})-E(n_{c},n_{0},n_{1})$,
$c^{\dag}c\otimes b_{1}^{\dag}
|n_{c},n_{0},n_{1}\rangle=n_{c}\sqrt{n_{1}+1}|n_{c},n_{0},n_{1}+1\rangle$,
$\omega_{1}=E(n_{c},n_{0},n_{1}+1)-E(n_{c},n_{0},n_{1})$.
$h_{s}(t)=\sum_{\omega}h_{s}(\omega)e^{-i\omega t}$,
$h_{int}(\tau)=h_{s}(\tau)\otimes A(\tau)$.

The bath correlations
(two-sided/symmetry correlation function) can be decomposed to the emission and absorption parts
\begin{equation}
	\begin{aligned}
		\mathcal{C}_{11}(\tau)
&=\int^{\infty}_{-\infty}g(\omega)e^{-i\omega\tau}d\tau={\rm Tr}_{b}[A(\tau)A\rho_{b}]
+{\rm Tr}_{b}[AA(\tau)\rho_{b}]\\
&
=\int^{\infty}_{0}g(\omega)(n(\omega)+1)e^{i\omega\tau}+
\int^{\infty}_{0}g(\omega)n(\omega)e^{-i\omega\tau}
=\frac{1}{2\tau_{b}}e^{-|\tau|/\tau_{b}}.
	\end{aligned}
\end{equation}
where ${\rm Tr}_{b}[A(\tau)A\rho_{b}]
={\rm Tr}_{b}[AA(\tau)\rho_{b}]^{*}
={\rm Tr}_{b}[AA(-\tau)\rho_{b}]
={\rm Tr}_{b}[AA(\tau-\frac{i}{T})\rho_{b}]
={\rm Tr}_{b}[A(-\tau+\frac{i}{T})A\rho_{b}]
=e^{\omega/T}{\rm Tr}_{b}[A(-\tau)A\rho_{b}]$ base on Kubo-Martin-Schwinger (KMS) relation,
and the last equality is base on the Lorentzian spectral density.
		
The corresponding Redfield equation reads
\begin{equation}
	\begin{aligned}
&\frac{d}{dt}\rho(t)
=-i[h_{s},\rho(t)]+\int^{\infty}_{0}
({\rm Tr}_{b}[h_{int}(t-\tau)\rho(t),h_{int}^{\dag}(t)]
+h.c.)d\tau\\
&=-i[h_{s},\rho(t)]+\int^{\infty}_{0}
({\rm Tr}_{b}[A(t-\tau)A(t)\rho_{b}]
[h_{s}(t-\tau)\rho(t),h_{s}^{\dag}(t)]
+h.c.)d\tau\\
&=-i[h_{s},\rho(t)]\\
&+\int^{\infty}_{0}
{\rm Tr}_{b}[A(t-\tau)A(t)\rho_{b}]
(h_{s}(t-\tau)\rho(t)h_{s}^{\dag}(t)
-h^{\dag}_{s}(t)h_{s}(t-\tau)\rho(t))\\
&+{\rm Tr}_{b}[A(t)A(t-\tau)\rho_{b}]
(h_{s}(t)\rho(t)h^{\dag}_{s}(t-\tau)
-\rho(t)h_{s}^{\dag}(t-\tau)h_{s}(t))d\tau
	\end{aligned}
\end{equation}
where the integral limit upto infinity and the density $\rho(t)$ independent of the $\tau$-integral imply the Markov approximation, and the bath memory time is much shorter than the system's evolution.
The above Redfield equation can be simplified to
\begin{equation}
	\begin{aligned}
&\frac{d}{dt}\rho
=-i[h_{s},\rho]\\
&+\int^{\infty}_{0}
{\rm Tr}_{b}[A(-\tau)A\rho_{b}]
(h_{s}(-\tau)\rho h_{s}^{\dag}
-h^{\dag}_{s} h_{s}(-\tau)\rho)
+{\rm Tr}_{b}[AA(-\tau)\rho_{b}]
(h_{s}\rho h_{s}^{\dag}(-\tau)
-\rho h^{\dag}_{s}(-\tau)h_{s})d\tau\\
&
=-i[h_{s},\rho]\\
&+\sum_{\omega}\int^{\infty}_{0}
{\rm Tr}_{b}[A(-\tau)A\rho_{b}]
(h_{s}(\omega)e^{-i\omega\tau}\rho h_{s}^{\dag}
-h^{\dag}_{s}h_{s}(\omega)e^{-i\omega\tau}\rho)\\
&+{\rm Tr}_{b}[AA(-\tau)\rho_{b}]
(h_{s}\rho h_{s}^{\dag}(\omega)e^{i\omega\tau}
-\rho h_{s}^{\dag}(\omega)e^{i\omega\tau}h_{s})
d\tau\\
&=-i[h_{s},\rho]
+\sum_{\omega}
\Gamma(-\omega)(h_{s}(\omega)\rho h_{s}^{\dag}
-h^{\dag}_{s}h_{s}(\omega)\rho)
+\Gamma^{*}(-\omega)
(h_{s}\rho h_{s}^{\dag}(\omega)
-\rho h_{s}^{\dag}(\omega)h_{s}),
	\end{aligned}
\end{equation}
where 
\begin{equation}
	\begin{aligned}
	&	h_{s}(\tau)=\sum_{mn}
|m\rangle\langle m|e^{iH_{s}\tau}h_{s}e^{-iH_{s}\tau}|n\rangle\langle n|\\
&=\sum_{mn}
|m\rangle\langle m|
e^{iH_{s}\tau}h_{s}(\omega)e^{-iH_{s}\tau}
|n\rangle\langle n|
=\sum_{mn}
e^{i(E_{m}-E_{n})\tau}
\langle m|h_{s}(\omega)|n\rangle
|m\rangle\langle n|\\
&=\sum_{\omega}
h_{s}(\omega)e^{i\omega\tau}.
	\end{aligned}
\end{equation}
The dissipation rate is the combination of spectral density $\gamma(\omega)$ and Lamb shift $\eta(\omega)$,
\begin{equation}
	\begin{aligned}
&\Gamma(\omega)
=\frac{1}{2}(\gamma(\omega)-i\eta(\omega))
=\int^{\infty}_{0}{\rm Tr}[A(\tau)A\rho_{b}]e^{i\omega\tau}d\tau,\\
&\gamma(\omega)=\int^{\infty}_{-\infty}{\rm Tr}[A(\tau)A\rho_{b}]e^{i\omega\tau}d\tau
=\Gamma(\omega)+\Gamma^{*}(\omega)=2{\rm Re}\Gamma(\omega),\\
&\eta(\omega)=i\int^{\infty}_{-\infty}{\rm sgn}\tau\ {\rm Tr}[A(\tau)A\rho_{b}]e^{i\omega\tau}d\tau
=i\Gamma(\omega)-i\Gamma^{*}(\omega)=-2{\rm Im}\Gamma(\omega),
	\end{aligned}
\end{equation}
which satisfies the KMS relation $\gamma(\omega)=e^{\omega/T}\gamma(-\omega)$,
$\eta(\omega)=-e^{\omega/T}\eta(-\omega)$,
and $\Gamma(\omega)=e^{\omega/T}\Gamma^{*}(-\omega)$.
$\Gamma(\omega)
=\int^{0}_{-\infty}{\rm Tr}[A(\tau)A\rho_{b}]e^{i\omega\tau}d\tau$.
Specifically, it satisfies the Hermitian symmetry $\gamma(\omega)=\gamma^{*}(-\omega)$
and anti-Hermitian symmetry $\eta(\omega)=-\eta^{*}(-\omega)$ when 
the bath correlation ${\rm Tr}_{b}[A(\tau)A\rho_{b}]$ is purely real
(${\rm Tr}_{b}[A(\tau)A\rho_{b}]={\rm Tr}_{b}[A(-\tau)A\rho_{b}]$)
in which case the real (imaginary) part of $\gamma(\omega)$ is even (odd) function of $\omega$,
and $\eta(\omega)=\eta^{*}(-\omega)$
($\gamma(\omega)=-\gamma^{*}(-\omega)$) when 
the bath correlation ${\rm Tr}_{b}[A(\tau)A\rho_{b}]$ is purely imaginary (${\rm Tr}_{b}[A(\tau)A\rho_{b}]=-{\rm Tr}_{b}[A(-\tau)A\rho_{b}]$).
The complex dissipation rate can be rewritten through the Kramers-Kronig relation as
\begin{equation}
	\begin{aligned}
		&\Gamma(\omega)
=\frac{\pi}{2} g(\omega)(n(\omega)+1)
+\frac{i\pi}{2}\mathcal{P}\int^{\infty}_{-\infty}\frac{g(\Omega)}{\omega-\Omega}(n(\Omega)+1)d\Omega.
	\end{aligned}
\end{equation}
The above trace-preserving Redfield equation is a time-local approximation and retains an oscillating and non-positive part (negative populations; due to mixing off-diagonal elements in a non-secular way that couples the population and coherence terms),
under the conditions of weak coupling and short bath correlation time (memory time).
The reduced density matrix $\rho(t)$ is local in time and is not time-evolved during the $\tau$ integration,
but the the bath correlation ${\rm Tr}[A(-\tau)A\rho_{b}]$ is 
non-local in time.

For Gibbs thermal steady state $\rho=e^{-n_{b}\omega/T}$ and under detailed balance condition, we have $[b^{\dag}b,b]=-b$,
$n_{b}b=b(n_{b}-1)$,
$\rho c^{\dag}c\otimes b_{0}=e^{-n_{b}\omega/T}c^{\dag}c\otimes b_{0}
=c^{\dag}c\otimes b_{0}e^{-(n_{b}-1)\omega/T}
=e^{\omega/T}c^{\dag}c\otimes b_{0}\rho$.
$[b^{\dag}b,b^{\dag}]=b^{\dag}$,
$n_{b}b^{\dag}=b^{\dag}(n_{b}+1)$,
$\rho c^{\dag}c\otimes b^{\dag}_{0}=e^{-n_{b}\omega/T}c^{\dag}c\otimes b_{0}^{\dag}
=c^{\dag}c\otimes b^{\dag}_{0}e^{-(n_{b}+1)\omega/T}
=e^{-\omega/T}c^{\dag}c\otimes b^{\dag}_{0}\rho$.
\begin{equation}
	\begin{aligned}
		&\frac{d}{dt}\rho
=-i[h_{s},\rho]
+\sum_{\omega}
\Gamma(-\omega)(h_{s}(\omega)\rho h_{s}^{\dag}
-h^{\dag}_{s}h_{s}(\omega)\rho)
+\Gamma^{*}(-\omega)
(h_{s}\rho h_{s}^{\dag}(\omega)
-\rho h_{s}^{\dag}(\omega)h_{s})\\
&=-i[h_{s}+H_{LS},\rho]
+
{\rm Re}\Gamma(0)(h_{s}\rho h_{s}^{\dag}
-\frac{1}{2}\{h^{\dag}_{s}h_{s},\rho\})
+{\rm Re}\Gamma^{*}(0)
(h_{s}\rho h_{s}^{\dag}
-\frac{1}{2}\{\rho,h_{s}^{\dag}h_{s}\})\\
\end{aligned}
\end{equation}
where $H_{LS}$ is the Lamb shift Hamiltonian which containing the component of ${\rm Im}\Gamma(0)$ and the non-Secular terms (finite $\omega$)
\begin{equation}
	\begin{aligned}
	H_{LS}	=\sum_{\omega}(
{\rm Im}\Gamma(-\omega)[h_{s}^{\dag},h_{s}(\omega)]
+{\rm Im}\Gamma^{*}(-\omega)[h_{s}^{\dag}(\omega),h_{s}])
=\sum_{\omega}
{\rm Im}\Gamma(-\omega)
([h_{s}^{\dag},h_{s}(\omega)]
-[h_{s}^{\dag}(\omega),h_{s}])
\end{aligned}
\end{equation}
The secular approximation (rotating wave approximation or Davies approximation) can be adopted by setting $\omega=0$ here,
and guarantees the complete positivity during the evolution of $\rho(t)$.
In this case, large Bohr frequency cause the fast the coherence oscillation
and only the non-oscillating diagonal terms survive
under the coarse-graining of system dynamics over a time interval much larger than the oscillation period.

The secular approximation is equivalent to performing a time-averaging (or coarse-graining) of the Redfield equation over a time interval $T_{cg}\gg\omega^{-1}$,
and the integral over fast-oscillating terms reads
\begin{equation}
	\begin{aligned}
	|	\frac{1}{T_{cg}}\int^{T_{cg}}_{0}e^{i\omega t}dt|=|\frac{e^{i\omega T_{cg}}-1}{i\omega T_{cg}}|\le \frac{2}{\omega T_{cg}},
\end{aligned}
\end{equation}
which vanish for large $\omega$ and equals 1 for $\omega=0$.

\section{Conclusion}

We consider a four-level electronic-bosonic system coupled to both a local mode and a continuous bosonic bath, provides a rich platform for studying the interplay between coherent dynamics, non-adiabatic couplings, and multi-channel dissipation. We consider two distinct theoretical frameworks, the trace-preserving Lindblad master equation and the non-Hermitian effective Hamiltonian (EHH) approach. 
The Lindblad equation describes the unconditional ensemble average, while the EHH describes the conditional dynamics of a no-jump trajectory. We established the quantitative boundaries (in time, temperature, and coupling strength) where the simpler EHH approximation accurately captures the system dynamics.
We also consider the zero-phonon line (ZPL) and phonon sideband (PSB) transitions, and provides a framework to study their modification by non-local dissipation.
The survival probability decay rate, $G(t)$, exhibits fundamentally different behaviors in the two frameworks, reflecting the difference between unconditional relaxation (Lindblad) and conditional survival. 
We characterized the decay dynamics, identifying regimes of standard exponential decay ($G(t) \sim t$) versus non-Markovian sub-exponential decay ($G(t) \sim t^\alpha, \alpha < 1$), linking the latter to Zeno-like suppression of thermalization.
We found that the stability and relaxation dynamics, as characterized by the Liouvillian gap, are starkly different. In the full Lindblad model, the gap vanishes at resonance ($\Delta\omega \to 0$), indicating critical slowing down. In contrast, the EHH gap remains finite, dominated by the non-Hermitian-related parameter $\text{Im}(W)$, demonstrating that the no-jump decay is robust against variations in coherent parameters.
The non-secular (rapidly oscillating) terms responsible for coupling coherences to populations, are involved for transitions between bosonic operators.
We apply the secular approximation
and rotating wave approximation for the transitions between the four levels,
where the system's evolution is much slower than the characteristic time scale of bath fluctuations (guaranteed by the markovianity such that the bath correlation time is instantaneous and without memory effects) and the external coherent driving is at the resonance.
Due to the existence of $H_{int;1}$
which is the interaction of non-adiabatic type, there is correlation between system and bath, and the non-markovianity can be observed in Hermitian case.

The results highlight the distinct physical information contained within conditional and unconditional open system dynamics. 
The introduction of a complex Bogoliubov parameter, $W$, as a way to model effective gain, loss, or phase effects in the mode transformation, opens a promising avenue for engineering and controlling non-Hermitian dynamics in finite-level quantum systems, with potential applications in quantum information processing and quantum thermodynamics.
Our results are helpful in
bridging the physics of molecular transitions with the formalisms of quantum optics and non-Hermitian dynamics.

		\clearpage
		
		\renewcommand\refname{References}
		
		\clearpage
		\section{Appendix.A: Deduction of Eq.(4)}
		
Next we present the deduction of
$\frac{W}{4}(b_{0}^{\dag}+b_{0})^2 = \sqrt{1+W}(b^{\dag}_{1}b_{1}+\frac{1}{2}) - (b^{\dag}_{0}b_{0}+\frac{1}{2})$.

\begin{equation} 
	\begin{aligned}
		&\sqrt{1+W}(b^{\dag}_{1}b_{1}+\frac{1}{2}) - (b^{\dag}_{0}b_{0}+\frac{1}{2})\\
		&=\sqrt{1+W}\left[ \frac{1}{4\sqrt{1+W}} \left((4+2W)b_{0}^{\dag}b_{0} + W((b_{0}^{\dag})^2+b_{0}^2) + (\sqrt{1+W}-1)^2\right) + \frac{1}{2} \right] - b^{\dag}_{0}b_{0} - \frac{1}{2}\\
		&= \frac{1}{4}\left((4+2W)b_{0}^{\dag}b_{0} + W((b_{0}^{\dag})^2+b_{0}^2) + (\sqrt{1+W}-1)^2\right) + \frac{\sqrt{1+W}}{2} - b^{\dag}_{0}b_{0} - \frac{1}{2}\\
		&= \frac{1}{4}\left((4+2W)b_{0}^{\dag}b_{0} + W((b_{0}^{\dag})^2+b_{0}^2) + (2+W-2\sqrt{1+W})\right) + \frac{\sqrt{1+W}}{2} - b^{\dag}_{0}b_{0} - \frac{1}{2}\\
		&=(1+\frac{W}{2})b_{0}^{\dag}b_{0} + \frac{W}{4}((b_{0}^{\dag})^2+b_{0}^2) + (\frac{1}{2}+\frac{W}{4}-\frac{\sqrt{1+W}}{2}) + \frac{\sqrt{1+W}}{2} - b^{\dag}_{0}b_{0} - \frac{1}{2}\\
		& = \frac{W}{2}b_{0}^{\dag}b_{0} + \frac{W}{4}((b_{0}^{\dag})^2+b_{0}^2) + \frac{W}{4}\\
		& = \frac{W}{4} \left( (b_{0}^{\dag})^2 + b_{0}^2 + 2b_{0}^{\dag}b_{0} + 1 \right)
		=\frac{W}{4}(b_0^\dagger + b_0)^2 
		=\frac{W}{4}(b_{0}^{\dag}+b_{0})^2,
	\end{aligned}
\end{equation}
where we use $(\sqrt{1+W}-1)^2 = (1+W) - 2\sqrt{1+W} + 1 = 2+W-2\sqrt{1+W}$,
$(b_0^\dagger + b_0)^2 = (b_0^\dagger)^2 + b_0^2 + b_0^\dagger b_0 + b_0 b_0^\dagger = (b_0^\dagger)^2 + b_0^2 + 2b_0^\dagger b_0 + 1$.
		
		\section{Appendix.B: Projected ensemble and "measurement" effect from the environment}
		
		At short time,
		the slower-than-ballistic (sub-ballistic) behavior of survival probability decay rate reflect the quantum Zeno effect where frequent and continuous interaction with the environment (or effetive "measurements") "freeze" a quantum system's evolution. The act of observation or coupling projects the system back into its initial state (or a small subspace), preventing it from evolving significantly away from it. This repeated "resetting" suppresses the natural unitary dynamics, and the effective "measurements" from environment keeping the system pinned to its initial state.
		The projected ensemble is a (incoherent) statistical mixture of the post-measurement states in
		the environment. Since each individual state in the ensemble is a product state, 
		the projective measurement suppress the entanglement between the system and environment
		by collapsing the entire entangled state (despite the classical correlations may remain).
		Meanwhile,
		this is also a process of unraveling\cite{Chang},
		which is by expressing this mixed state of environment
		as a classical probability distribution over a set of pure states. 
		Since the density matrix itself contains all the information about the system, 
		this ensemble of pure states can be used to reproduce any measurement outcome that would be obtained from the mixed state. 
		For unreplicated Hilbert space (first moment), the reduced density matrix
		\begin{equation} 
			\begin{aligned}
				\rho_{E}={\rm Tr}_{S}\rho_{SE}=\sum_{z}p_{z}|\psi_{z}\rangle\langle\psi_{z}|
				=\sum_{i}\lambda_{i}|\psi_{i}\rangle\langle\psi_{i}|,
				\label{1}
			\end{aligned}
		\end{equation} 
		where the pure states $|\psi_{i}\rangle$ are orthonormal eigenvectors of the $\rho_{E}$,
		$\lambda_{i}$ are the eigenvalues (real and non-negativ) of $\rho_{E}$.
		$p_{z}$ are the classical probabilities of the measurement outcome on the system labelled by index $z$.
		The full scrambling as a result of standard thermalization 
		retains only the averages constrained by global conservation and loss the microscopic details carried by the pure states
		(e.g., initial pure state's fluctuation). 
		The statistical ensemble averaging 
		over the disordered degrees of freedom of the system 
		approaches the maximally mixed state and gives rise to
		the typicality of the type of canonical (Gibbs) or GGE (Scrooge ensemble), i.e., $\rho_{E}$ becomes statistically indistinguishable with ${\rm Tr}_{S}\rho_{eq}$
		even when the total system is in a pure state.
		Here, a thermal equilibrium state is a mixed state that maximizes entropy subject to global conserved quantities. 
		It's highly disordered and has a large amount of classical correlations and statistical mixedness.
		The thermal states at finite temperature is a statistical mixture where states with lower energy 
		are more likely to be occupied (biased distribution of pure states), 
		and thus distinct from the unbiased Haar ensemble (whose average state is the maximally mixed state).
		For higher moments, it reads
		\begin{equation} 
			\begin{aligned}
				\rho_{E}^{(k)}=\sum_{z}p_{z}(|\psi_{z}\rangle\langle\psi_{z}|)^{\otimes k}.
			\end{aligned}
		\end{equation} 
		Different to the above standard thermalization,
		the Haar ensemble represents a uniform probability distribution over the pure states of
		the entire Hilbert space whose unbiased average leads to maximally mixed state.
		deep thermalization: it looks at the entire distribution of possible post-measurement states, not just their average. 
		Here replacing the reduced density matrix by the
		classical statistical mixture of pure states,
		captures information about the classical correlations between the post-measurement states
		with different measurement outcome $z$ and different $k$.
		Then the specific post-measurement states (conditioned projected ensemble)
		behaves away from maximally mixedness and classical ignorance.
		The property of typicality shows that,
		a certain type of
		probability distribution in Hilbert space feature similar quantum expectation values of generic observables\cite{Bartsch}.
		
	As a consequences of a zero Liouvillian gap, there would be more than one non-steady-state eigenvectors of the Liouvillian with eigenvalues having a real part of zero. This leads to multiple steady states.
This is related to the non-defective degeneracy\cite{wu} where the zero-eigenvalue subspace is larger than one-dimensional,
and the eigenstates in a zero-eigenvalue subspace are orthogonal to each other.
The long-time behavior depends on the initial state's projection onto the entire nullspace instead of a single state, and the final state is a superposition of these steady states.

	\end{small}

\begin{thebibliography}{99}
\bibitem{squee}
The operators $X_i = \frac{b_i^\dagger + b_i}{\sqrt{2}}$ and $P_i = \frac{b_i^\dagger - b_i}{i\sqrt{2}}$ are the quadrature operators (generalized position and momentum) that define the uncertainty relation and squeezing.
The ratio $\frac{(b_{1}^{\dag}+b_{1})^2}{(b_{0}^{\dag}+b_{0})^2} = \sqrt{1+W}{\bf I}$ is the square of the quadrature ratio.In squeezing, the variance of one quadrature is reduced (e.g., $\Delta X_1 < \Delta X_0$), and the other is increased ($\Delta P_1 > \Delta P_0$). This transformation is often parametrized by a squeezing parameter $r$, where $X_1 = e^{-r} X_0$ and $P_1 = e^{r} P_0$. 
In this case, the ratio $\frac{X_1}{X_0}$ directly relates to the squeezing factor.
Here, it can also be defined as the standard single-mode squeeze operator $S(\zeta) = \exp(\frac{1}{2}\zeta^* b_0^2 - \frac{1}{2}\zeta (b_0^\dagger)^2)$ where $\zeta = r e^{i\phi}$ is the complex squeezing parameter and the Bogoliubov transformation $b_1 = S(\zeta) b_0 S^\dagger(\zeta)$ is a related mathematical structure, where the parameters $u, v$ can be written as $u=\cosh|\zeta|, v=e^{i\arg(\zeta)}\sinh|\zeta|$. 
Specifically, for real $W$, $\sqrt{1+W} = e^{2r}$ links $W$ to the squeezing amplitude $r$.
\end{thebibliography}
\end{document}